\documentclass[sigplan]{acmart}
\renewcommand\footnotetextcopyrightpermission[1]{}
\AtBeginDocument{%
  \providecommand\BibTeX{{%
    \normalfont B\kern-0.5em{\scshape i\kern-0.25em b}\kern-0.8em\TeX}}}

\setcopyright{acmcopyright}
\copyrightyear{2018}
\acmYear{2018}
\acmDOI{XXXXXXX.XXXXXXX}

\acmPrice{15.00}
\acmISBN{978-1-4503-XXXX-X/18/06}

\usepackage{url}
\usepackage{graphicx}
\usepackage{latexsym}
\usepackage{amsfonts}
\usepackage{comment}
\usepackage{array}
\usepackage{xspace}
\usepackage{psfrag}
\usepackage{color}
\usepackage{enumitem}
\usepackage{alltt}
\usepackage{multirow}
\usepackage{array}
\usepackage{wasysym}
\usepackage{rotating}
\usepackage{datetime}
\usepackage{amsthm}
\usepackage{epsfig}
\usepackage{endnotes}
\usepackage{mathrsfs}
\usepackage{epstopdf}
\usepackage{booktabs}
\usepackage{enumerate}

\usepackage{subcaption}

\usepackage{indentfirst}
\usepackage[noend]{algpseudocode}

\usepackage{tikz}

\usepackage{algorithmicx, algorithm}

\usepackage{cleveref}

\newcommand{\ie}{\textit{i.e.}\xspace}
\newcommand{\eg}{\textit{e.g.}\xspace}

\newcommand{\sys}{{ProNet}\xspace}

\newcommand{\presub}{\vspace{-0.1in}}
\newcommand{\postsub}{\vspace{-0.05in}}

\begin{document}

\title{\sys: Network-level Bandwidth Sharing among Tenants in Cloud}


\author{Zhewen Yang}
\affiliation{
  \institution{Johns Hopkins University}
    \city{Baltimore}
  \country{USA}
}
\email{zyang122@jh.edu}

\author{Changrong Wu}
\affiliation{
  \institution{Nanjing University}
      \city{Nanjing}
  \country{China}
}

\author{Chen Tian}
\affiliation{
  \institution{Nanjing University}
        \city{Nanjing}
  \country{China}
}

\author{Zhaochen Zhang}
\affiliation{
  \institution{Nanjing University}
        \city{Nanjing}
  \country{China}
}







\renewcommand{\shortauthors}{Trovato and Tobin, et al.}

\begin{abstract}
In today’s private cloud, the resource of the datacenter is shared by multiple tenants. 
Unlike the storage and computing resources, it's challenging to allocate bandwidth resources among tenants in private datacenter networks.
State-of-the-art approaches are not effective or practical enough to meet tenants' bandwidth requirements.
%
%
%
%
%
%
%
In this paper, we propose \sys, a practical end-host-based solution for bandwidth sharing among tenants to meet their various demands.
%
%
%
The key idea of \sys~is byte-counter, a mechanism to 
collect the bandwidth usage of tenants on end-hosts 
%
to guide the adjustment of the whole network allocation, without putting much pressure on switches. 
%
%
We evaluate \sys~both in our testbed and large-scale simulations.
Results show that \sys~can support multiple allocation policies such as network proportionality and minimum bandwidth guarantee.
Accordingly, the application-level performance is improved.
\end{abstract}

\maketitle

\section{Introduction}
\label{sec1}
Cloud resources (e.g., storage, computing, and network) are shared among multiple tenants according to their individual requirements and payments.
In resource management in cloud, a Service Level Agreement (SLA) is usually set between the Cloud Provider and the cloud user to meet the requirements of the allocation of some resources in terms of the performance (like availability, resource amount and cost). 
All these SLAs are mainly set to guarantee the Quality of Service (QoS) in the cloud. The QoS metric is vary widely according to the context (like edge computing, cloud mobile, IoT). 
Most SLAs can successfully measure and optimize the physical resource in computing and storage like CPU, Memory, and disk. 
In the same time, network resource also plays an essential role in the QoS. 
The network condition decide the network latency, transmitting efficiency and quality due to the bandwidth of each service in the cloud. 
Therefore its management is important to satisfy the bandwidth requirements of tenants and achieving good application performance in datacenter networks.
%

%
%

However, network bandwidth allocation is different from storage and computing resources allocation.
Unlike storage and computing resources that can be allocated at a fixed ratio, network resources are usually shared among tenants dynamically.
It is difficult to guarantee the bandwidth allocation via a simple static end-to-end reservation.
A bandwidth allocation approach should meet several design requirements as follows~\cite{ballani2011towards,ballani2013chatty,xie2012only,angel2014end,lee2014application,guo2010secondnet,lam2012netshare,jang2015silo}.
Firstly, multiple allocation policies should be supported simultaneously.
The bandwidth demands of tenants can be various, including minimum guarantee (this is what most previous allocation systems try to achieve), proportional allocation, and even according to specific utility functions (\S~\ref{sec:background:allocation}, \ie a more complex allocation strategy than the former two).
Secondly, in datacenters, bandwidth is usually shared by different tenants (workgroups) with different requirements.
Hence, high bandwidth utilization, \ie, work conservation, is required to save the cost.
%
%
Besides, the application-level performance should not be compromised when ensuring bandwidth allocation.
Last but not least, the approach should be deploy-friendly and suitable to the cloud environment.
State-of-the-art approaches are not effective or practical enough to allocate bandwidth according to tenants' requirements in cloud environment.
%
%
%
Most of the bandwidth allocation approaches only target ensuring a minimum bandwidth guarantee~\cite{jeyakumar2013eyeq,popa2013elasticswitch,chowdhury2016hug,hu2018providing}. And we think only achieving the minimum guarantee of bandwidth among tenants is not enough to fit the complex cloud environment and some specific SLA's requirements in nowadays cloud. Besides, most of them can't achieve work conservation, so the network resources are wasted at the most time. 
%
For the proportionality allocation, google designed BwE for there WAN environment. However, the environment of the implementation of BwE~\cite{kumar2015bwe} is a SDN scenario, all routing information and the flow transmitting plan is known by the manager. This is impossible for the cloud environment so the BwE can't be used in datacenters nowadays.
A natural way to achieve proportionality allocation is to leverage in-network priority queues to ensure weighted fair queuing and isolation.
HCSFQ~\cite{yu2021twenty} can approximate weighted priority queues, but it can not meet other requirements among tenants.
In addition, its active packet drop behavior can harm the application-level performance of the network.
%
PS-N~\cite{popa2012faircloud} aims at providing proportionality allocation by leveraging a weight model to distribute weights among flows.
It delivers an ideal proportional allocation model.
%
%
%
Unfortunately, PS-N requires at least as many weighted fair-sharing queues as the number of tenants (\eg, several thousand) on each switches, which is impractical.
In addition, PS-N is limited to take effects in specific scenarios, \ie, identical (or proportional) link capacities and tenant weights.
%
%
%


%
%

%
To this end, we propose \sys, a practical end-host-based network bandwidth allocation protocol without burdening switches.
%
%
In our opinion, the main feature of the bandwidth inside a cloud environment is considered as limited in local areas, but unlimited in the whole picture.  
Therefore, the goal of \sys is not to achieve the instance bandwidth allocation fairness among tenants, but to keep the right allocation strategy in a period of time. 
The key design of \sys~is \textit{byte-counter}, a simple yet effective mechanism to dynamically adjust the bandwidth allocation among tenants based on their real usage without knowing the real networking situation inside the cloud.
%
%
The intuition of bytes-counter is that the usage of a tenant can be measured by counting the volume of transmitted data, therefore by collection the bandwidth usage of each tenants, the usage can be converted to the volume of data transmitted in the corresponding period of time.
If the total number of transmitted bytes can be controlled appropriately at end-hosts, the bandwidth over the whole network can be allocated correctly.
Byte-counter does not make assumptions on bandwidth or targets of tenants, so it is flexible to be applied to a wide range of scenarios.
By leveraging byte-counter, \sys~does not rely on switches to support a large number of queues.
\sys~can meet multiple allocation policies in a deploy-friendly way (\S~\ref{sec:key-idea:bytes-counter}).
%
%

%
%
To meet all requirements mentioned above, \sys~proposes several design points aside from byte-counter.
%
%
%
%
To satisfy utility bandwidth functions of tenants, \sys~leverages bandwidth functions to allocate bandwidth hierarchically (\S~\ref{sec:key-idea:hierarchical}).
It also makes \sys~able to provide a flexible bandwidth allocation.
%
%
%
%
%
%
To achieve work conservation, \sys~proposes Congestion-Aware Work-Conserving mechanism (CAWC) to perceive the network congestion state on receiver end-hosts, guiding senders to adjust the transmission rate of tenants, and converge the allocation strategy into a conserving state in a short time.
In addition, tenant-counter is leveraged to differentiate intra-tenant congestion (\ie, congestion caused by intra-tenant flows) from inter-tenant congestion (\S~\ref{sec:system:rate-controller}).
We implement a prototype of \sys~using 10 servers also with Tofino1~\cite{tofino1} switches.
Testbed experiments and large-scale NS-3 \cite{ns3} simulations validate that \sys~can converge to the predefined bandwidth allocation targets.
%
%
%
%
Taking the ideal PS-N as the baseline, \sys can achieve the all properties mentioned above and converge in ms scale. Compared with HCSFQ~\cite{yu2021twenty}, \sys~achieves 29\% better average throughput and reduces the flow completion time (FCT) by 24\%, benefiting from reducing packet loss and improving the network utilization.
%
%


\section{Background and Motivation}\label{sec:background}
\label{sec2}
\setlength{\parskip}{0.5em}

\presub\subsection{Tenant-based Bandwidth Allocation}\postsub ~\label{sec:background:per-tenant}
%
%
%
%
In modern datecenters, cloud service providers give infrastructure services to tenants simultaneously through virtualization. 
Not only for the computing and storage resources, the bandwidth is an essential part of the network resource. The bandwidth of each tenant's tasks decide the performance and the quality of the transmitting, and whether it satisfy the SLAs.
Neither a flow-based nor a source-destination-based allocation is suitable for providing cloud services for tenants.
Also, traditional bandwidth allocation methods usually target distributing bandwidth for each flow.
%
%
%
A tenant can increase its bandwidth share by increasing the number of its flows maliciously without paying more.
%
%
%
Allocating bandwidth based on per source-destination pair can aggregate the bandwidth used by flows belonging to the same source-destination.
Likewise, it can not deal with the cases where a tenant increases the number of destinations it connects to, getting more bandwidth share than those not.
In order to provide bandwidth according to tenants' payments, a per-tenant bandwidth allocation approach is necessary.
Meanwhile, flows belonging to the same tenants can have different bandwidth demands.
Their individual demands should be satisfied at the same time do not violate the bandwidth allocation among tenants.

%
%

\presub\subsection{Properties Required}\postsub~\label{sec:background:allocation}
%
%
To meet the bandwidth demands of tenants and ensure high application-level performance, bandwidth allocation approaches should satisfy the following properties, which are preferred to industries.
%
%

\begin{itemize}
\item \textbf{Support Multiple Policies.}
There are usually multiple specific bandwidth demands for different tenants. 
%
%
A bandwidth allocation approach is necessary to satisfy multiple policies simultaneously by providing flexible run-time reconfiguration.
Here, we introduce some of the most vital specific properties for the network allocation among tenants.
%
%

%
\textit{Network Proportionality.} 
%
%
When sharing the bandwidth between tenants, bandwidth should be allocated proportionally based on their payments or priorities.
In other words, when two tenants are competing for the same network resource, the allocation of this resource should be according to some ratio or priority.

%
\textit{Minimum Bandwidth Guarantee.} 
The bandwidth allocated to tenants should at least equal the minimum bandwidth according to their demands. 
This property is essential for tasks that are sensitive to bandwidth or flow completion time\cite{ghemawat2003google,dean2008mapreduce}.

\textit{Utility bandwidth functions.} 
Tenants could specify their bandwidth demands in the form of utility bandwidth function.
A bandwidth allocation approach support bandwidth allocation according to tenants' utility bandwidth functions is more flexible.


\item \textbf{Work Conservation.}
It denotes that links in a datacenter (usually the congested links) are fully utilized or meet all demands. 
Being work conservation means an effective network and high bandwidth utilization.

\item \textbf{Application-level Performance.}
The application-level performance should not be compromised when providing the allocation policies~\cite{wilson2011better}.
Hence, the latency of flows should be reduced, at the same time ensuring high throughput and low packet loss rate.
%
%
%

\item \textbf{Deploy-friendly.}
An allocation approach that can be easily deployed is preferred to datacenter networks.
Hence, the requirements put on switches should be reduced.

\end{itemize} 

%


\begin{table*}[t]
	\setlength{\belowcaptionskip}{0mm}
	\setlength{\abovecaptionskip}{-3 mm}
\resizebox{\linewidth}{!}{
\begin{tabular}{|l|llll|ll|}
\hline
\textbf{}                        & \multicolumn{4}{l|}{\textbf{Design Properties}}                                                                                                                                & \multicolumn{2}{l|}{\textbf{System Requirments}}                                                                                                                                                  \\ \hline
\textbf{Prior Works}             & \multicolumn{1}{l|} {\textbf{\begin{tabular}[c]{@{}l@{}}Min Bandwidth\\ Guarantee\end{tabular}}} & \multicolumn{1}{l|}{\textbf{Proportionality}} & \multicolumn{1}{l|} {\textbf{\begin{tabular}[c]{@{}l@{}}Work\\ Conservation\end{tabular}}} & \multicolumn{1}{l|} {\textbf{\begin{tabular}[c]{@{}l@{}}Low Latency\\ and High Throughput\end{tabular}}} & \multicolumn{1}{l|}{\textbf{Hardware Support}}                                                                                                                      & \textbf{Topo and flow situation Requirement} \\ \hline
\textbf{Oktopus\cite{ballani2011towards}}          & \multicolumn{1}{l|}{\checkmark}                              & \multicolumn{1}{l|}{$\times$}                         & \multicolumn{1}{l|}{$\times$}                       & $\times$                   & \multicolumn{1}{l|}{None}                                                                                                                                             & None                      \\ \hline
\textbf{Seawall\cite{shieh2010seawall}}           & \multicolumn{1}{l|}{$\times$}                               & \multicolumn{1}{l|}{$\times$}                        & \multicolumn{1}{l|}{\checkmark}                       & $\times$                   & \multicolumn{1}{l|}{None}                                                                                                                                             & None                      \\ \hline
\textbf{EyeQ\cite{jeyakumar2013eyeq}}       & \multicolumn{1}{l|}{\checkmark}                              & \multicolumn{1}{l|}{$\times$}                        & \multicolumn{1}{l|}{\checkmark}                       & $\times$                   & \multicolumn{1}{l|}{None}                                                                                                                                             & Congestion-free core      \\ \hline
\textbf{FairCloud PS-P\cite{popa2012faircloud}}   & \multicolumn{1}{l|}{\checkmark}                              & \multicolumn{1}{l|}{$\times$}                        & \multicolumn{1}{l|}{\checkmark}                       & $\times$                   & \multicolumn{1}{l|}{None}                                                                                                                                             & Tree                      \\ \hline
\textbf{ElasticSwitch\cite{popa2013elasticswitch}}    & \multicolumn{1}{l|}{\checkmark}                              & \multicolumn{1}{l|}{$\times$}                        & \multicolumn{1}{l|}{\checkmark}                       & $\times$                   & \multicolumn{1}{l|}{None}                                                                                                                                             & None                      \\ \hline
\textbf{FairCloud PS-L/N \cite{popa2012faircloud}} & \multicolumn{1}{l|}{$\times$}                               & \multicolumn{1}{l|}{\checkmark}                        & \multicolumn{1}{l|}{\checkmark}                      & $\times$                   & \multicolumn{1}{l|}{Per-VM queues on switch}                                                                                                                                    & None                      \\ \hline
\textbf{BwE \cite{kumar2015bwe}} & \multicolumn{1}{l|}{\checkmark}                               & \multicolumn{1}{l|}{\checkmark}                        & \multicolumn{1}{l|}{/}                      & /                  & \multicolumn{1}{l|}{None}                                                                                                                                    & Topo, flow infos and allocation plans                      \\ \hline
\textbf{Trinity\cite{hu2018providing}
}           & \multicolumn{1}{l|}{\checkmark}                              & \multicolumn{1}{l|}{$\times$}                        & \multicolumn{1}{l|}{\checkmark}                       & \checkmark                  & \multicolumn{1}{l|}{Priority queues on switch}                                                                                                                                  & None                      \\ \hline
\textbf{HCSFQ\cite{yu2021twenty}}             & \multicolumn{1}{l|}{/}                                & \multicolumn{1}{l|}{\checkmark}                        & \multicolumn{1}{l|}{/}                      & $\times$                    & \multicolumn{1}{l|}{\multirow{2}{*}{\begin{tabular}[c]{@{}l@{}} Approximate queue model\\  on programmable switches.\end{tabular}}} & /                         \\ \cline{1-5} \cline{7-7} 
\multicolumn{1}{|l|}{\textbf{AIFO\cite{yu2021programmable}}}              & \multicolumn{1}{l|}{/}                                & \multicolumn{1}{l|}{$\times$}                        & \multicolumn{1}{l|}{/}                      & \multicolumn{1}{l|}{$\times$}                   & \multicolumn{1}{l|}{}                                                                                                                                                 & \multicolumn{1}{l|}{/}                         \\ \hline
\textbf{\textbf{GearBox}\cite{gao2022gearbox}}              & \multicolumn{1}{l|}{/}                                & \multicolumn{1}{l|}{\checkmark}                        & \multicolumn{1}{l|}{/}                      & $\times$                    & \multicolumn{1}{l|}{FPGAs on smart NIC. }                                                                                                                                                 & /                         \\ \hline
\textbf{\sys~}                     & \multicolumn{1}{l|}{\checkmark}                              & \multicolumn{1}{l|}{\checkmark}                        & \multicolumn{1}{l|}{\checkmark}                      & \checkmark                  & \multicolumn{1}{l|}{None or Little}                                                                                                                                   & None                      \\ \hline
\end{tabular}
}
\caption{Properties and requirements comparison of state-of-the-art approaches.}
\label{tab1}
\end{table*}
	
\presub \subsection{Related Works} \postsub
%
The sharing of cloud infrastructure inevitably causes widespread competition on resources, among which network bandwidth allocation is one of the most critical and challenging issues. 
%
%
%
Almost all state-of-the-art bandwidth allocation approaches cannot achieve all of the properties simultaneously with a practical and deployable design.
We first introduce the approaches to solving the minimum bandwidth guarantee for tenants (or flows). 
Generally, these protocols seek spare bandwidth in the network for tenants with inadequate allocation or simply limit the bandwidth for the over-used flows at end-hosts.
%
EyeQ~\cite{jeyakumar2013eyeq} provides tenants with bandwidth guarantees. However, it is based on a non-blocking switch model, which does not handle in-network congestion.
Oktopus~\cite{ballani2011towards} provides predictable bandwidth guarantees while it does not achieve work conservation, only resulting in low network efficiency.
Besides, it is not scalable.
ElasticSwitch\cite{popa2013elasticswitch} relies on end-to-end rate control and can not achieve fine-grained bandwidth allocation.
%
%
%
PS-P~\cite{popa2012faircloud} achieves the bandwidth guarantee across tenants perfectly with the help of weighted fair queuing on switches.

Meanwhile, other approaches focus on different types of fairness in bandwidth allocation. 
These approaches achieve fairness by leveraging rate monitors and transmission control on switches.
%
%
Seawall\cite{shieh2010seawall} provides per-source fairness in congested links.
However, this is useless for multi-tenant scenarios. 
NUMFabric\cite{nagaraj2016numfabric} provides a flexible and configurable allocation framework that can achieve different allocation targets, such as weighted allocation. 
However, it puts much pressure on the programmable switch for the algorithm calculation jobs, and WFQ is also required in some particular jobs. 
%
FairCloud~\cite{popa2012faircloud} has proposed two methods of weighted bandwidth allocation: PS-L targets at link proportionality and PS-N targets at congestion proportionality. 
The link proportionality of PS-L means keeping the allocation fairness among each sender-receiver link pear, which is unnecessary for cloud environments.
These two proposals require as many queues as the number of tenants, which is impractical (\S~\ref{sec:background:motiv}).
What's more, these three methods in FairCloud can only achieve some specific allocation weights and plans instead of a flexible and more complex allocation strategy.
BwE\cite{kumar2015bwe} is a bandwidth allocation system developed by Google, and it can achieve the proportional allocation with a flexible and hierarchical straucture. However, BwE is designed for Google WAN, which is a SDN-like environment. In Google's WAN, all tasks plans and the flow information are known and set before the tasks by the administrator. Also the in-time flow status can also be awarded by the network managers. And BwE is designed in this strong prerequisites, which is a totally different environment in cloud. Therefore BwE can't solve the allocation job inside the datacenter. 
%
%
%
%

Table~\ref{tab1} summarizes the state-of-the-art bandwidth allocation approaches and compares them according to whether they satisfy the properties mentioned in~\S~\ref{sec:background:allocation} and their requirements on switches and network topologies.

\presub \subsection{Motivation} \postsub \label{sec:background:motiv}
%
%
In private enterprise networks, different types of bandwidth demands should be met (\S~\ref{sec:background:allocation}).
The most central goal among them is providing proportionality bandwidth allocation over the whole network, \ie, weighted fair share.
Since cloud providers should ensure proportional services according to tenants' payments, which is also called \textit{fairness} among tenants.
%
%
%
%

%
\noindent\textbf{Weight fair queuing on switches is not practical and yet not enough.} 
A natural way to satisfy proportional bandwidth allocation is to provide one weighted fair queue (WFQ)~\cite{demers1989analysis} for each tenant on switches.
However, it is infeasible for each tenant to require one queue on switches for WFQ.
The number of tenants can be orders of magnitude more than the number of queues that a switch can support (\eg, at most 32/128 queues per port in the latest programmable switches~\cite{tofino2}). 
Many queuing scheduling designs have appeared in recent years using relatively limited resources on programmable switches and other newly developed programmable devices. 
AIFO~\cite{yu2021programmable} tries to use a single FIFO queue to achieve priority queue, but it can hardly achieve the weighted bandwidth fairness allocation. 
Gearbox~\cite{gao2022gearbox} tries to approximate the WFQ by using a hierarchical FIFO-based scheduler. 
However, its process is pretty complex and is not able to be applied to this generation's programmable switch. 
Instead, Gearbox is implied on the smart NICs, which can support FPGAs.
The WFQ only applied on hosts' NICs is obviously not enough for the bandwidth allocation of the network environment in a datacenter. 
Also, the scheduling time scale of Gearbox is also limited by the number of the FIFO queues. 
The most practical WFQ design is HCSFQ~\cite{yu2021twenty}. 
However, in its design, HCSFQ needs to drop packets proactively until achieving fairness which could cause a large amount of retransmission.
%
%
%
%
%
It can downgrade the application-level performance and is usually not expected, especially for tasks aiming at low latency.
Moreover, solutions to approximate WFQs are unable to support other tenants' requirements, such as minimum guarantee.
%

%
\noindent\textbf{End-host-based approaches to meet bandwidth allocation demands are challenging.} 
%
%
Controlling bandwidth allocation on end-hosts straightforwardly does not take in-network congestion into account.
However, the condition of datacenter networks is complex and varies over time.
In-network congestion is not uncommon, \eg, the network topology can be oversubscribed, and incast traffic occurs intermittently~\cite{alizadeh2011data}.
A flow can encounter congestion and compete for bandwidth with other traffic in the network.
Hence, the actual bandwidth used by the flow does not equal the bandwidth allocated on the end hosts.
%
%
%
%
%

\noindent\textbf{The bandwidth allocation in cloud is hard as well as achieving the work conservation.}
In the cloud environment, it is impossible to predict
or get to know the flow information head of the allocation
job. There are multiple load balances in the datacenter like
ECMP. The flow in the datacenter is dynamic and have large real time variations. The only information of a task or flow the manager
can know is the src and the dst of it, instead of knowing the
actual links and path of each flow like in WAN or LAN. So
that the allocation job should be done without knowing the
allocation plan and even the topology in the cloud like BwE.
Also, it is important to achieve the work conservation. This means the bottleneck of each tasks should be used without waste. Work conservation also means the high utilization of the bandwidth, and the bandwidth is usually linked with the latency and the QoS. However, the flow and task is unpredictable in cloud, so the the place of bottlenecks and the allocation situation in bottlenecks are also unknown and variable. So achieving the work conservation is also a target to overcome. 

\noindent\textbf{Applicable scenarios are limited.} 
%
%
%
%
%
%
%
The state-of-the-art approach aiming at a proportional fair share is a bandwidth allocation model called Proportional Sharing at Network-level (PS-N), which is proposed in FairCloud\cite{popa2012faircloud}. %
PS-N relies on per-tenant WFQs.
PS-N can only achieve network-level fairness in a restricted condition where all bottleneck links have the same capacity and background weight, or all congested links are proportionally loaded. 
This is impractical. 
Datacenter network is a complex system with many random factors, \eg, unequal links and transient bursty traffic. 
%
%

%
%
%

%
\noindent\textbf{Unfeasible to meet variable tenants' goals.} 
The demands of tenants in clouds can change along with their application traffic. 
Cloud providers should cater to the tenants' demands in real-time.
However, due to the constant calculation pattern of PS-N, it can only keep a simply fixed allocation ratio between flows, which can not satisfy the variable demands of tenants.
%
%

\begin{figure*}[t]
  \centering
	\setlength{\belowcaptionskip}{-2mm}
	\setlength{\abovecaptionskip}{4 mm}
  \resizebox{\textwidth}{!}{
    \begin{subfigure}[t]{0.26\textwidth}
      \centering
      \resizebox{\textwidth}{!}{
        \includegraphics{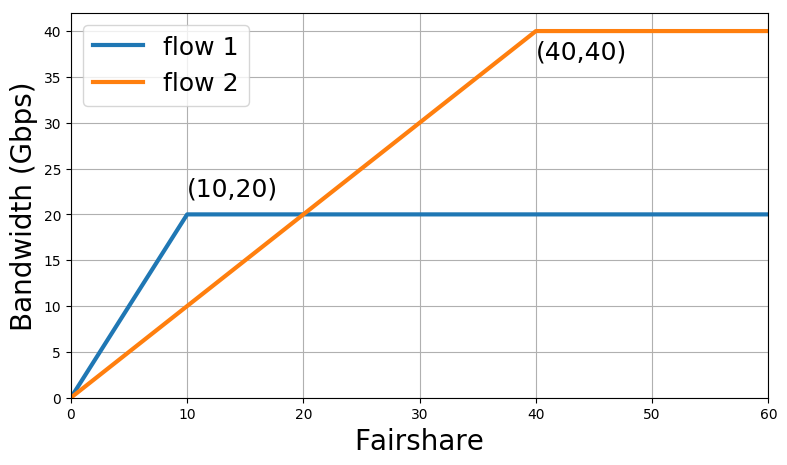}
      }\captionsetup{font=scriptsize}
      \caption{Bandwidth functions of flows.}
      \label{fig1a}
    \end{subfigure}\hfil
    
    \begin{subfigure}[t]{0.26\textwidth}
      \centering
      \resizebox{\textwidth}{!}{
        \includegraphics{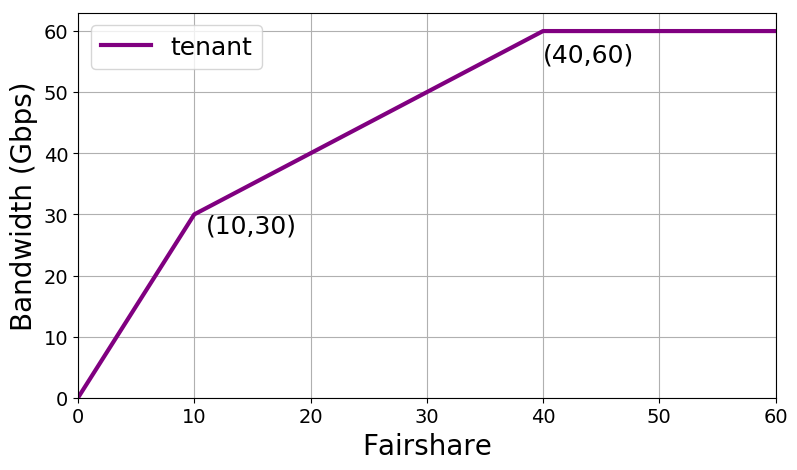}
      }\captionsetup{font=scriptsize}
      \caption{Bandwidth function of the tenant.}
      \label{fig1b}
    \end{subfigure}\hfil

    \begin{subfigure}[t]{0.26\textwidth}
      \centering
      \resizebox{\textwidth}{!}{
        \includegraphics{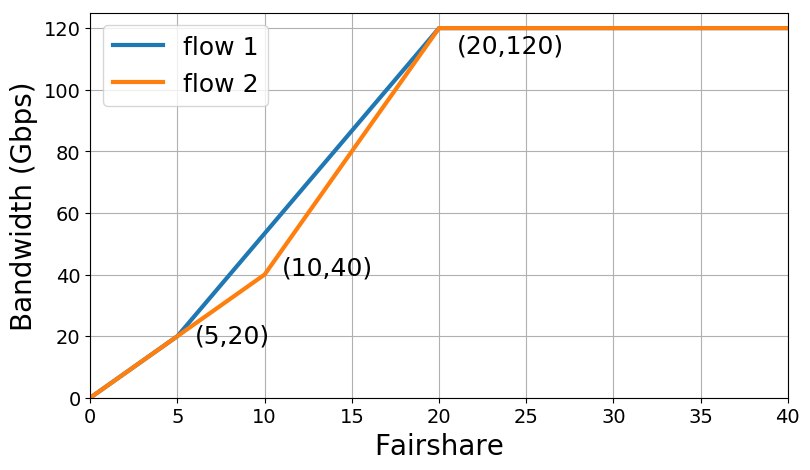}
      }\captionsetup{font=scriptsize}
      \caption{Flows' BFs after aggregation}
      \label{fig1c}
    \end{subfigure}\hfil

  }
  \caption{An example to illustrate how flows' bandwidth functions (BFs) are aggregated with their tenants'.}
  \label{fig:bf}
\end{figure*}

\presub \subsection{Bandwidth Function} \postsub
Bandwidth Function (BF) has been used in Google's Bandwidth Enforcer (BwE) system in their private WAN \cite{jain2013b4, kumar2015bwe}.
BF specifies the bandwidth allocation to a flow group as a function mapping from a dimensionless available bandwidth share capacity measurement, called \textit{fair share}, to the actual bandwidth allocation value.
%
%
It is feasible and effective to model the variable allocation demands of tenants into BFs.
Compared to fixed-weighted fair sharing, BFs can represent more complex demands of tenants and flows.
In fact, the fixed weighted fair sharing and minimum guarantee can be denoted by the simplest BFs.
%
%
Moreover, when the tenants' demands change, BFs can be re-configured conveniently, without requesting changes or reconfiguration in networks.
Therefore, \sys~leverages BF to support run-time reconfiguration and is flexible to bandwidth demands.
\section{Key Ideas}\label{sec:key-idea}
To achieve the properties discussed in \S~\ref{sec:background:allocation}, we propose \sys. 
In this section, we introduce several key design points of \sys, including byte-counter and hierarchical control of tenants.


%
\subsection{Byte-counter}~\label{sec:key-idea:bytes-counter}
%
One of the most challenging parts of an allocation system is to adapt to the bandwidth demands without relying on switches to support multiple queues.
%
%
%
%
\sys~leverages byte-counter to estimate the bandwidth usage of each tenant on end-hosts.
As the name byte-counter implies, it records the number of bytes sent by each tenant in a period of time.
\sys's end-hosts maintain a byte-counter for each tenant to record the sent bytes.
Since flows of a tenant can be distributed on multiple end-hosts, it is necessary to aggregate byte-counter on different end-hosts for each tenant.
Hence, the local byte-counter is reported to the coordinator periodically.
%
%
%
%
%
%
%
By leveraging byte-counter for each tenant, the average bandwidth occupation during the corresponding period of time can be calculated.
The average bandwidth can help \sys~to adjust the bandwidth allocation of tenants in the next cycle.
There can be scenarios where the bandwidth reports from hosts are delayed, and it does not affect the convergence of \sys~for that the error of each cycle does not accumulate (\S~\ref{coordinator}).

Byte-counter helps \sys~to strike a balance between accurate bandwidth allocation and practical.
It is overlooked by most prior works that tenants target achieving overall good application-level performance instead of requiring instant bandwidth proportionality.
Instead of focusing on the instantaneous bandwidth usage of the sending ports (queues), byte-counter aims at achieving network proportionality over the network in the long term.
%
%
Although \sys~can not converge to the target weighted fair share instantly,
\sys~does not compromise the performance of small flows since the bytes used by small flows can be counted accurately.
%

%
%
%
\begin{figure}
	\centering
	\setlength{\belowcaptionskip}{-5mm}
	\setlength{\abovecaptionskip}{1 mm}
	\includegraphics[width=1\linewidth]{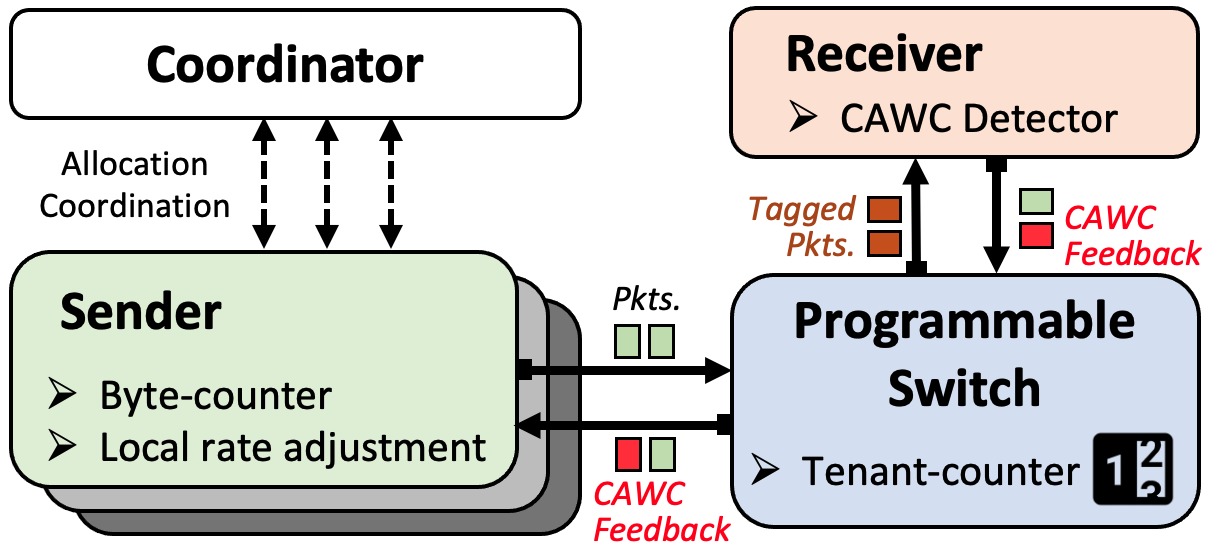}
	\caption{System Structure.} 
	\label{fig2}
\end{figure}

\begin{figure*}[t]
  \centering
\setlength{\belowcaptionskip}{-1mm}
	\setlength{\abovecaptionskip}{2 mm}
  \resizebox{\textwidth}{!}{
    \begin{subfigure}[t]{0.23\textwidth}
      \centering
      \resizebox{\textwidth}{!}{
        \includegraphics{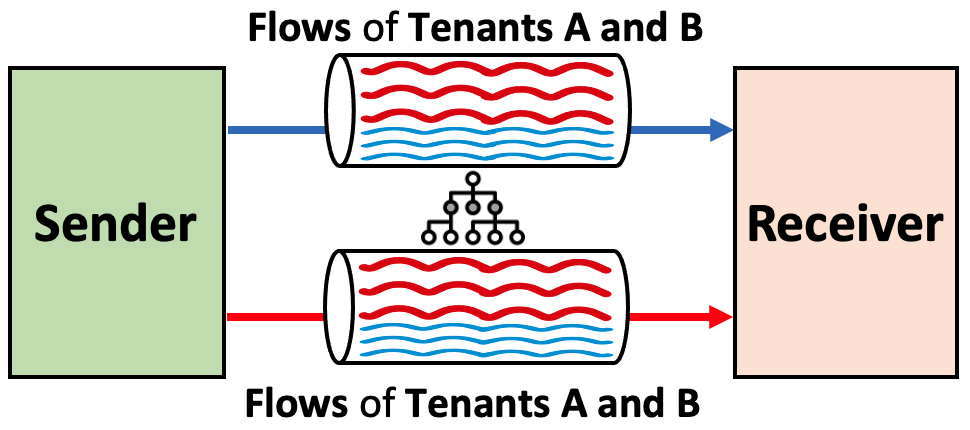}
      }
      \captionsetup{font=scriptsize}\caption{Normally inter-tenant congestion}
      \label{cona}
    \end{subfigure}\hfil
    
    \begin{subfigure}[t]{0.23\textwidth}
      \centering
      \resizebox{\textwidth}{!}{
        \includegraphics{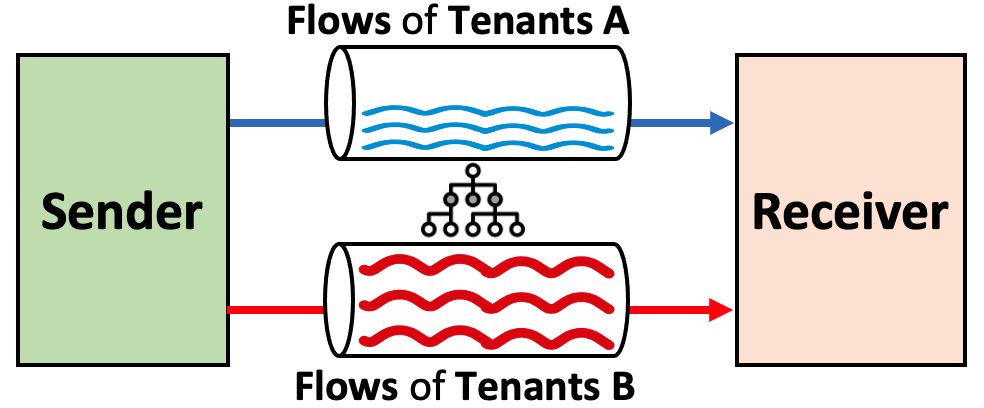}
      }
      \captionsetup{font=scriptsize}\caption{Intra-tenant congestion with a \protect\\ poor network utilization}
      \label{conb}
    \end{subfigure}\hfil

    \begin{subfigure}[t]{0.23\textwidth}
      \centering
      \resizebox{\textwidth}{!}{
        \includegraphics{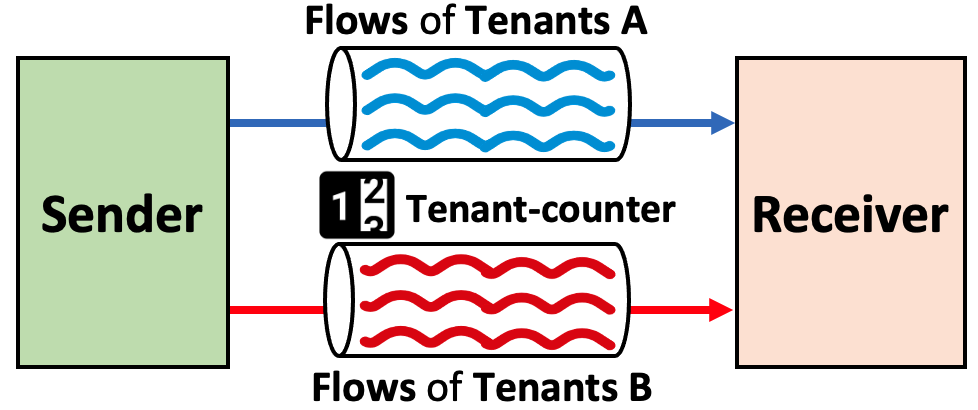}
      }\captionsetup{font=scriptsize}
      \caption{Fixed intra-tenant congestion with tenant-counter}
      \label{conc}
    \end{subfigure}
  }
  \caption{An example to illustrate different congestion conditions among tenants, \ie, inter-tenant and intra-tenant congestion. 
  The target weight ratio among tenants A and B is 2:1.}
  \label{congestion}
\end{figure*}
\subsection{Hierarchical Control of Tenants}\noindent~\label{sec:key-idea:hierarchical}
%
%
\hspace{-0.08in}\noindent\textbf{Unit-flow.}
%
%
%
Applications with different scheduling objectives and bandwidth demands require isolation. 
In order to achieve the fine-grained management of flows belonging to the same tenant, we create the \textit{unit-flow} abstraction. 
Unit-flow is a group of packets sharing the same source and destination pair with the same tenant ID.
Unit-flow is the minimum control unit of data in \sys. 
%
%
%
%
%

\noindent\textbf{Bandwidth function aggregation.} 
To meet the varied and customized demands of tenants, bandwidth functions (BF) is leveraged to configure the sharing policy.
Each tenant uses its own bandwidth functions, and so as the flows (unit-flows) belonging to it.
%
%
%
%
\sys~allocates bandwidth in a hierarchical way, \ie, bandwidth is allocated to tenants according to their BFs, and then the bandwidth is shared among the tenant's flows according to flows' BFs. 
In order to satisfy the bandwidth demands of tenants and flows simultaneously, bandwidth functions of flows should be able to represent the requirements of its belonged tenants.
Inspired by the hierarchical MultiPath Fair Allocation (MPFA) algorithm \cite{kumar2015bwe} targeting WAN distributed computing, \sys~aggregates bandwidth functions of a flow with its belonged tenants' BF (\S~\ref{tbfa}).
%
%
%
Lower level flows' BFs are transferred into aggregated BFs with a higher level tenant's BF as input by leveraging aggregation functions.
%
%

%
In Figure~\ref{fig:bf}, we take a quick look at how flows' bandwidth functions are aggregated with their tenants' BF.
%
There are two flows, and Figure~\ref{fig1a} shows their bandwidth functions.
%
%
%
%
Figure~\ref{fig1b} shows the BF of the tenant to which the two flows belong, which represents a total of different allocation requirements. 
Figure~\ref{fig1c} shows the BFs of flows after aggregation.
New features from their tenant have been acquired. The specific algorithm is shown in \S~\ref{tbfa}. 
%
%
%
%

%
%
%

%
%
%
%
%
%
%
%
%
%


\subsection{End-host Based congestion detection and work conservation accomplish}\noindent~\label{sec:key-idea:End}
%
The most difficult challenge for an end-host-based network system is how to detect the in-network situation on hosts.
%
%
%
To detect in-network congestion among tenants ant to achieve a more efficient work conservation goal, \sys's end-hosts leverage the Congestion-Aware Work-Conserving mechanism (CAWC) to detect in-network congestion.
CAWC is responsible for removing uncongested flows from the control loop of the byte-counter to improve network utilization.
%
%
%
%

%
Besides, there are corner cases where congestion is only induced by flows belonging to the same tenant, \ie, \textbf{intra-tenant} flows.
In this paper, we identify network behaviors within a tenant as \textbf{intra-tenant}, and those among tenants as \textbf{inter-tenant}. 
\sys~should leave the rate adaptation of those flows to congestion control protocols to avoid bandwidth waste.
Figure \ref{congestion} shows the dilemma.
The weight ratio between tenants A and B is 1:2, and the two links are identical.
%
Figure \ref{cona} shows the normal cases where both two links have traffic from both tenants.
With \sys, tenant A's flows get 2/3 of the total bandwidth, and tenant B's flows get the rest.
But when tenants do not compete for bandwidth, things are different.
Figure \ref{conb} shows the scenario where each tenant's flows occupy a link separately. 
%
%
Due to the proportional bandwidth allocation, flows of tenant A only get half of the bandwidth, which results in the bandwidth waste of the upper link.
%
%
In fact, tenants A and B should not be considered to be allocated in a weighted manner since they do not compete for bandwidth. 
%
%
\sys~leverages tenant-counter to record the number of tenants passing through the links, which can be easily implemented in programmable switches (\S~\ref{sec:system:tenant-counter}).
%
%

\section{System Design}~\label{sec:design}
\label{sec4}
%

%
%
Built from the key ideas (\S~\ref{sec:key-idea}), we propose \sys, a host-based practical network bandwidth allocation protocol for tenants.
Figure \ref{fig2} shows the basic framework of \sys. 
\sys is composed of three parts: sender, coordinator, and receiver.
%
%
When a new flow arrives, the traffic controller of the sender groups them into unit-flows and initiates their states (\S~\ref{sec:system:pre-processing}).
The sender's rate controller is responsible for the bandwidth allocation.
%
\sys~allocates bandwidth hierarchically.
The bandwidth of tenants is allocated according to the fair share calculated by the coordinator and their bandwidth functions.
Then, the bandwidth of flows within a tenant is allocated accordingly (\S~\ref{sec:system:traffic-controller}).
In addition, to guide the rate adaptation of flows and achieve work conservation, the network congestion state should be perceived (\S~\ref{sec:system:rate-controller}).
\subsection{Allocation Behavior Initiation}\noindent~\label{sec:system:pre-processing}
%
%
%
%
%
%

%
When new flows arrive, the traffic controller regroups the flows and groups them into unit-flows.
Each unit-flow is assigned a bandwidth function.
%
%
%
Generally, the flows' bandwidth function is initialized according to the weights of the source and destination pair and the rate limiter set on the sender hosts.
Equation \eqref{1} denotes the bandwidth function of flow $x$, where $srcWeight$ and $dstWeight$ denote the weight of the source host and the destination host, and $deviceRateLimit$ denotes the rate limit on the sender host. This BF takes in the fairshare $s$ as the function's input. 
\begin{equation}\label{1}
B_x(s)=\min{\left(srcWeight+dstWeight,\ deviceRateLimit\right)} 
\end{equation}
%

%
%
%
This bandwidth function represents the preference among different source and destination pairs.
In fact, the bandwidth function supports customized initialization.
%
%
By setting different bandwidth functions to the unit-flows, \sys~can obtain flexible and versatile bandwidth allocation. 
For instance, a tenant could have preferences among different sender hosts, \eg, one tenant may need one host to obtain twice as much bandwidth as others.
This can be achieved by simply initializing the bandwidth function to the minimum value of $srcWeight$ and $deviceRateLimit$.
In addition, a minimum bandwidth guarantee can be achieved by simply initializing the bandwidth functions to the bandwidth guarantee.
Also, more complex BFs for unit-flows to achieve more individualized allocation demands are also capable.
%

%
%
%

%
\subsection{Bandwidth Allocation}\noindent~\label{sec:system:traffic-controller}
%
%
%
%
%
%
%
%
In this subsection, we first introduce the bandwidth function aggregation of tenants (\S~\ref{tbfa}), which is fundamental to hierarchical bandwidth allocation (\S~\ref{sec:system:bandwidth-allocation:hierarchical}).
%
%

%
\subsubsection{Tenants' Bandwidth Function Aggregation}\label{tbfa}
%
%
%
%

%
Generally, a bandwidth function (BF) is a sectional incremental continuous function so that it can be represented as a group of interesting points (the points whose gradient of the function is zero) conveniently. 
Also, it is easy for a BF to get its inverse function. 
This means taking a bandwidth value as the input, and we can get the unique corresponding fair share with the BF.
These properties can be leveraged to process bandwidth function aggregation.

The major process is shown in \ref{appendix:a}. In this way, \sys~can get aggregated BFs for each unit-flow which can both satisfy the BF of the unit-flow and the tenant it belongs to. That's essential for \sys to coordinate between tenants and their network flows.
Due to BFs can be stored as sets of points in \sys, this algorithm can be easily applied.
As shown in Figure \ref{fig:bf}, the interesting points in the original BFs of flows 1 and 2 are transformed into the points shown in Figure \ref{fig1c} with the BF of the tenant in Figure \ref{fig1b}.

\subsubsection{Hierarchical Bandwidth Allocation}\label{sec:system:bandwidth-allocation:hierarchical}
\begin{figure}
	\centering
	\setlength{\belowcaptionskip}{-5mm}
	\setlength{\abovecaptionskip}{2 mm}
	\includegraphics[width=3in]{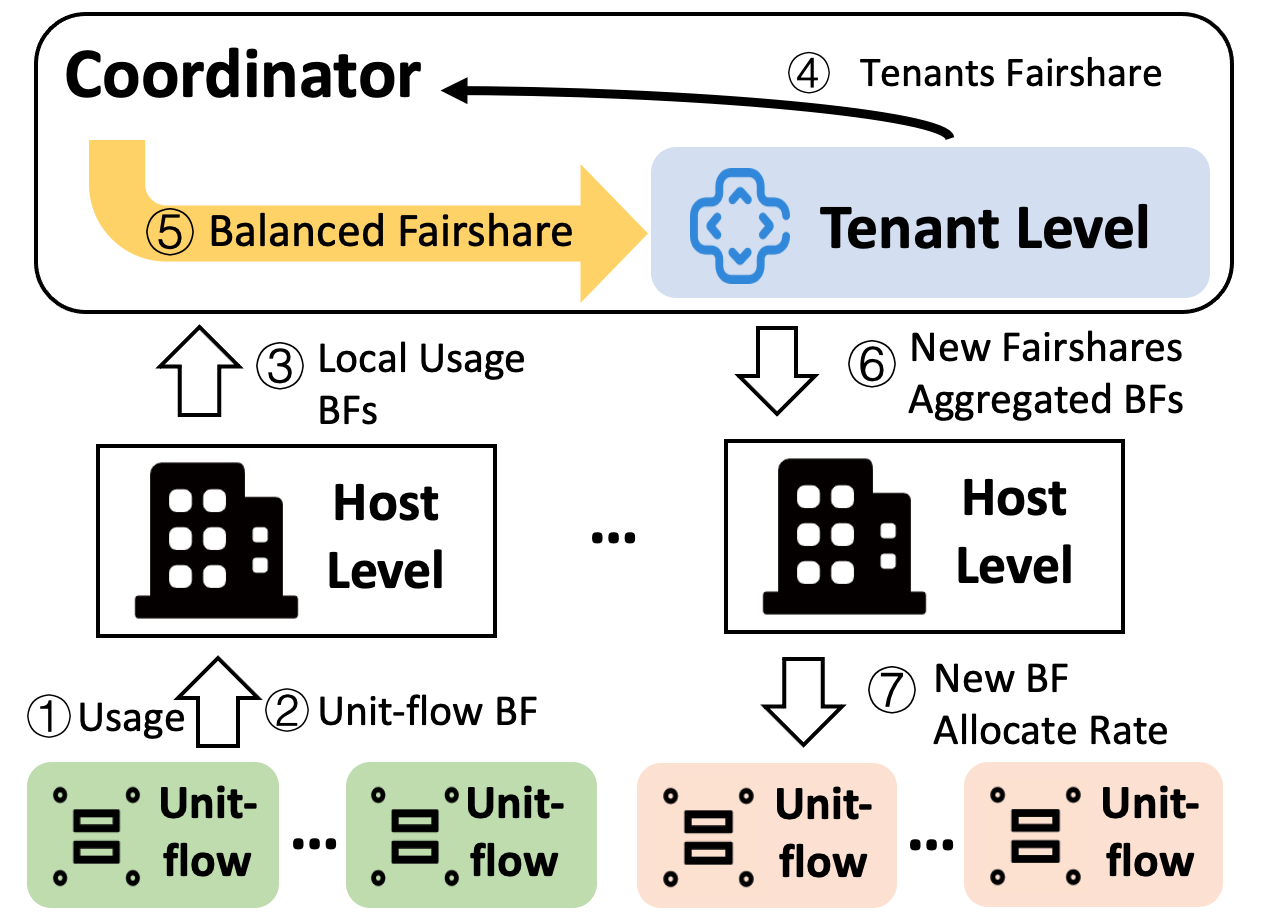}
	\caption{Information Flow Process}\label{fig3}
\end{figure}

\sys~controls the bandwidth allocation between tenants and their flows in a hierarchical way, \ie, from flows to their tenant, then to the whole network level.
The fair share and the bandwidth functions are calculated hierarchically.
The bandwidth usage of flows belonging to the same tenants is firstly aggregated to calculate a local fair share.
Then, the local share is calculated by leveraging the coordinator to interact with all end-hosts in networks.
Likewise, the bandwidth function aggregation works in a similar way.
\noindent\textbf{Intra-tenant bandwidth allocation.}
The rate controller adjusts the sending rate of each flow by leveraging whole network information exchanged from the coordinator. 
Figure \ref{fig3} shows the architecture of the rate controller.
%
%
The rate controller calculates the fair share of local tenants by taking the usage of unit-flows into account and exchanges the fair share with the coordinator to get synchronized network states.
%
%
%

%
We now discuss how \sys~allocates bandwidth within a single tenant by leveraging tenant controllers at the tenant level. 
At the start, tenant controllers are assigned with the initial bandwidth function (\S~\ref{sec:system:pre-processing}), which reflects the allocation policy of the tenant. 
For each cycle time, the total bandwidth usage is collected by using byte-counter, as well as the bandwidth functions of unit-flows belonging to the corresponding tenant. 
After that, the tenant controller aggregates all these bandwidth functions with the tenant's bandwidth function, getting new bandwidth functions for those unit-flows.
Therefore, all unit-flows in this tenant will get a brand new bandwidth function that satisfies both the allocation targets of its tenant and the unit-flow itself. 
It ensures that unit-flows of the belonged tenant can be assigned with the correct bandwidth, which is necessary to achieve further fairness allocation among the whole network.
After unit-flows are allocated with appropriate bandwidth, their transmission rate should be limited according to their obtained bandwidth.
Token Bucket Filter (TBF) \cite{glasmann2000estimation}is leveraged as the rate limiter for unit-flows. 
For that TBF can provide relatively stable rate control, and its peak rate is controllable. 
%
%
%
%
The token bucket increases its tokens according to the bandwidth allocated to the unit-flow.
Before a packet is transmitted, the traffic controller checks whether the token bucket contains sufficient tokens. 
If so, the packet is sent, and the number of tokens equivalent to the bytes of the packet is removed.
%
%
%
%

%
%
%
%

%
%
%
%
%
%
%
%

%

\noindent\textbf{Coordinator for inter-tenant bandwidth allocation.}\label{coordinator}
There are many end-hosts in the datacenter, \sys~should be able to coordinate all of them to achieve network level fairness. 
%
For each host, the fair share of every tenant should be kept at the same level all the time to ensure fairness. 
The coordinator is leveraged to interact with end-hosts. 
To avoid the coordinator becoming the network bottleneck, the logic of the coordinator is quite simple.
For each reporting cycle of each host, the coordinator collects the overall bandwidth usage (\ie, the fair share value) of each tenant from end-hosts' rate controller and then sends back the updated target fair share.
%
%
%

The mean update process come up with an $InputArray=[s_1,\ s_2,\ s_3,\ \ldots,\ s_N]$ which stands for the input array containing all the collected fair share from each tenant. $s_i$ denotes the fair share of tenant $i$.
After aggregating all these values, the coordinator will then calculate as $Avg=average(InputArray)$ to average the possible allocation unfairness condition of the last cycle. 
Besides, an accelerating factor $alpha$, is also added to all fair shares as: $Target\ =\ Avg\ \ast\ \left(1\ +\ alpha\right)$. 
After calculation, the coordinator sends back the updated value of the fair share to all the rate controllers.
%
%
After each host receives the new target's fair share from the coordinator, it then calculates the new bandwidth (rate) for each unit-flow using its bandwidth functions. 
And the updated rate is sent to the TBF for rate-limiting. 
The following process is similar to the intra-tenant bandwidth allocation.
And Figure \ref{fig3} shows the information flow progress of \sys.

Also, as for the coordinate pattern, due to the report time of each byte-counter being varied, it is necessary to support an asynchronous report and coordinate pattern. 
To achieve this goal, in our design, the coordinator keeps a report window for byte-counters to report their usages. Within a report window, the coordinator will balance all the usage of the reported unit-flows in the corresponding hosts. To avoid the potential allocation unfairness due to the reporting delay of each host, we set a rule that if a host-only continues the next cycle of balanced allocation when an allocation instruction from the coordinator is received, \ie, the host won't send its usage report twice with only one report window on the coordinator. 
\subsection{Local Rate Adjustment}~\label{sec:system:rate-controller}
%
%
%

%
%
The flows' rate can not be simply adjusted according to the received fair share. 
Instead, the rate adaptation should take the network congestion states into consideration.
%
%
There can be bursty intermittent flows that come and leave quickly.
Hence, flows should occupy a higher bandwidth when congestion does not occur to utilize the network.
At the same time, the rate of flows passing through the congestion paths should be reduced to avoid aggravating congestion in the network. 
%
%
%
Furthermore, to ensure work conservation, the intra-tenant congestion should be left for congestion control instead of leveraging \sys.
%
%
%
%

%

%

%
%
\subsubsection{CAWC Mechanism}~\label{sec:system:receiver}
%
\sys's receiver is responsible for detecting the in-network congestion state according to the congestion information carried in packets, called the Congestion-Aware Work-Conserving mechanism (CAWC).
%
%
CAWC is leveraged to detect whether the network bottleneck is fully loaded or not. In this way, to achieve the work conservation aspect of \sys.
%
%
\sys's receiver maintains a scoreboard to count the bytes of received packets within a given period of time $sliding\_time$ (\eg, 10 $\mu$s), according to whether they are marked with ECN or not.
%
%
%
When the ratio of ECN-marked packets exceeds a given threshold $congestion\_thre$, the receiver sends back the congestion signal notification to the sender.
%
%
Note that the CAWC signal uses the highest priority to guarantee in-time delivery.
When the sender receives the CAWC signal, it adjusts the flows' fair share according to the Algorithm~\ref{algo:distribute}.
%
%
%
The control loop of uncongested flows is separated from the congested flows, \ie, the transmission of uncongested flows is not affected by the in-network congestion that they do not contribute to.
This can avoid unnecessary bandwidth waste and ensure work conservation.
%
%
Besides, the transmission control of congested flows can be more robust to scenarios where bursty small flows leave the network quickly by setting the congestion ratio.
And the allocation will converge in a real short time, normally in a sub-ms scale.

\begin{algorithm}\footnotesize

 
%
\caption{Distributed Rate Adaptation Algorithm}
\hspace*{0.02in} {\bf Input:} 
\label{algo:distribute}
\newcommand{\Note}[1]{\hfill\textcolor[rgb]{.25,.25,.25}{\small // #1}}
$TargetFS$ \Note{updated fair share received from the coordinator}

\hspace*{0.02in} $k$ \Note{the rate attenuation factor when congested}

\hspace*{0.02in} $rateControlCycle$ \Note{the execution period of this algorithm}


\hspace*{0.02in} $reportCycle$ \Note{the communication interval to the coordinator}

\hspace*{0.02in} $oldFS$ \Note{the fair share of last cycle}\\
\hspace*{0.02in} $newFS$ \Note{the fair share of next cycle}
%
\begin{algorithmic}[1]
\State This algorithm runs at the $rateControlCycle$ periodically
\For{$flow$ in flowTable} 
	\If{$flow$ does not transmit in the last cycle} 
	\State Mark the flow as inactive
	\State \textbf{continue}
	\EndIf
	\If{$oldFS \neq TargetFS$}
	\State $nFS$ = the compensation fair share according to the BF of $flow$
	\State $oldFS$ += $nFS$
	\EndIf

	\If{congestion is detected} 
    \If{congestion is detected in the prior cycle}
	\State $newFS=oldFS-k$
	\Else
	\State $newFS=oldFS$
	\EndIf

	\Else
	\State $newFS=oldFS*(1+rateControlCycle/reportCycle)$
	\EndIf
	\State Calculate $nBW$ with $newFS$ as the input of the BF of $flow$
	\State $rateSum+=nBW$
\EndFor
\State  $scalingFactor = 1$
\If{$rateSum >= deviceRateLimit$}
\State  $scalingFactor *= (deviceRateLimit / rateSum)$
\EndIf
\For{active $flow$ in flowTable}
\State Set the allocated rate of $flow$ as $nBW *scalingFactor$
\EndFor
\end{algorithmic}
\end{algorithm}

\subsubsection{Distributed Rate Adaptation Algorithm}\label{sec:system:distribute-algorithm}
%
%
The main idea of \sys is that the bandwidth is infinite in the whole picture of cloud, but it's finite in local and instantly. So in this part we concentrate on the adjustment of bandwidth allocation in the granularity of tenant in a period of time.
The rate controller at the end-host runs a rate adaptation algorithm to optimize the bandwidth allocation distributedly, as demonstrated in Algorithm~\ref{algo:distribute}.
%
%
%
%
%
%
%
%
The host handles each unit-flow recorded in the flowTable to allocate their rates periodically. 
The unit-flows are checked whether it is active first. 
%
%
%
%
Only the active unit-flows are allocated with bandwidth (Line 2-5).
\sys~compares the updated fair share TargetFS received from the coordinator with the fair share oldFS of the flow in the last allocating cycle, where oldFS is initialized according to the starting rate of flows.
In order to keep the overall allocation fairness, \sys~calculates a new fair share for the unit-flow, which can compensate for the unfairness in the last cycle allocation.
The integration of the compensation nFS equals the integration of TargetFS-oldFS, \ie, $\int_{oldFS}^{oldFS+nFS} BF_f(s)\, ds = \int_{oldFS}^{TargetFS} BF_f(s)\, ds$ (Line 7-8).
Note that $nFS$ can be negative numbers when $oldFS$ is larger than $TargetFS$. 
After this, the host adjusts each unit-flow's rate according to the network state and the local bandwidth usage. 
To avoid being too sensitive to congestion and utilize the network, here we use two consecutive rate control cycles to determine whether a unit-flow takes part in the in-network congestion (Line 9-15).
If the congestion is detected for the first time, the allocation is maintained as the old fair share $oldFS$.
But for the second time, the bandwidth allocation for this unit-flow is decreased by k to alleviate the congestion.
If no congestion is detected, \ie, the flow does not take part in the congestion, its rate is increased by a fraction less than double its rate to climb up to utilize the bandwidth (Note that $rateControlCycle$ is smaller than the $reportCycle$).
The bandwidth allocation can be calculated by taking the fair share and the bandwidth function as input (Line 16). 
When the total allocation on the host reaches the rate limit, \sys~ scales the overall allocation (Line 19-20).  
Later, with the new calculated fair share, \sys~can then assign the next allocation target bandwidth of each unit-flow (Line 22). %
%

%
%

    


%
With this algorithm, every host can distributedly optimize the next cycle allocation goal for each unit-flow. 
%
%
For unit-flows that contribute to the in-network congestion are enforced for rate adaptation.
%
%
As for uncongested unit-flows are enforced to utilize the network to achieve work-conserving.
%
%
Evaluations show that the sending hosts can converge quickly with several rounds of rate adaptation (\S~\ref{subsec:pqt}).


%


%
%

\subsubsection{Tenant-counter}\label{sec:system:tenant-counter}
To add the ability of \sys to detect the difference between inter-tenant and intra-tenant congestion, we design an optional module called tenant-counter on the programmable switch.
What \sys~requires for the programmable switch is just a  \textit{Tenant register table}. 
%
%
This is a table with a preset capacity (we usually set as 2) of entries to record the tenant information of passing by flows.
In this table, we keep at most a number of (2) least met tenants' ID and recent encountered time.
When a packet (flow) passes through the switch, the table registers its corresponding tenant ID. 
When another tenant's packet passes this switch, a new entry is added to the table. 
%
%
To reduce the network overhead, the switch does not transmit the tenant table.
Instead, if the congestion occurs (\ie, exceeding the ECN-marking threshold) and the number of entries is larger than one, the packet is carried with an inter-tenant congestion flag directly.
The receiver carries back the flag to the sender by adding an extra bit in the congestion signals.
In addition, the expired tenants are removed from the table when the switch does not receive their flows for a timeout value which can be calculated with the recorded encounter time.
%
%
In addition, our evaluation shows that the tenant-counter is not necessary to be applied to all switches.
Instead, tenant-counter can be applied to the switch where congestion is more likely to occur, \ie, the last hop of the network or the oversubscribed nodes.
%
%
%

On the host-end, a module called non-competitive pool is added. 
At first, all unit-flows are set as non-competitive and classified into this pool. 
%
%
Unit-flows in the pool are not controlled by \sys.
%
%
After receiving the inter-congestion signals from the receiver, the corresponding unit-flows are moved from the pool for further allocation control.
%
%
%
The non-competitive pool is a useful abstraction that can be leveraged for user-defined performance optimization.
%
%
%
%
For instance, small flows can be put into the non-competitive pool for better performance.
If some flows require a high priority or should maintain the quality of service, they can also be assigned to the non-competitive pool.
%

\section{Implementation}
\label{sec6}

%

%
We implement a prototype of \sys in both real machine testbed and ns-3 simulation codes. 
We built \sys as a user-level process that implement the token bucket filter to rate-limit the transmission rate of flows in the Linux kernel. Our current implementation has around 2000 lines of C++ code.
To evaluate \sys, we build a dumbbell testbed with 16 servers connected to  Pronto-3295 48-port Gigabit switches and setup a fat-tree (k=3) topo to simulate the datacenter structure. 
We have also turned on the ECN and ECMP of our testbed to for the congestion control and the load banlancer to simulate the cloud environment.

\begin{figure}
	\centering
	\setlength{\belowcaptionskip}{-5mm}
	\setlength{\abovecaptionskip}{2 mm}
	\includegraphics[width=0.95\linewidth]{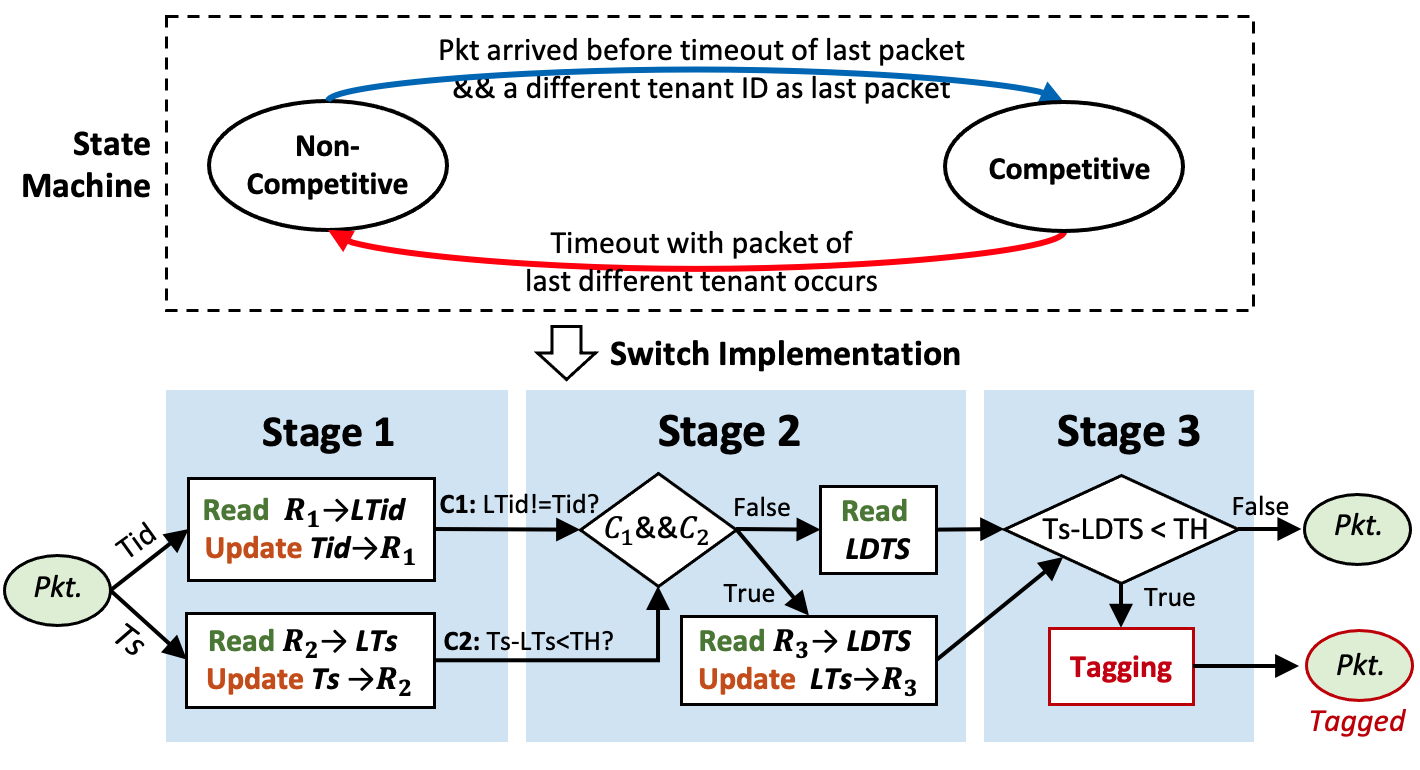}
	\caption{P4 Program State Machine and Procedure. (Tid: Tenant ID, Ts: Timestamp, LTid: Tenant ID of the last packet, LTs: Timestamp of last packet, LDTS: Timestamp of the last packet from a different tenant, TH: Timeout threshold. R1, R2, and R3 are stateful memory (\ie, registers) in the switch.)}\label{66}
\end{figure}

The tenant-counter is implemented on Tofino1~\cite{bosshart2014p4}.
Tenant-counter only requires little help from the programmable switch.
Three registers of the data plane are used.
The state machine of our P4 program logic is shown in Figure \ref{66}. 
There are two main states of an output port of \sys's switch: competitive and non-competitive.
Competitive denotes that there are multiple tenants' flows forwarded to the same output port.
Packets are supposed to be tagged to notify the end hosts.
A non-competitive state denotes that the output link is occupied by a single tenant's flows, and packets passing through a non-competitive port are forwarded as normal. 
The initial state of the switch's port is non-competitive. There are two conditions that should be satisfied for the state transformation. First, the arriving packet should belong to a different tenant from the prior packet's tenant. Second, the time interval between these two packets should be less than a preset timeout threshold. The above two conditions stand for the beginning of an inter-tenant congestion, and the state is transformed into a Competitive state. Next, the switch is supposed to tag the incoming packet for the notification for the receiver.
%
Also, the switch is supposed to be aware of whether the competition no longer appears in the output ports. Therefore, the state will be transferred to the initial state when this condition is satisfied: the time interval between the arriving packet and the last different tenant's packet is larger than the timeout threshold. This stands for the termination of an inter-tenant congestion.
%
%

To satisfy the state machine we mentioned above, we develop the P4 program of our testbed, also shown in Figure \ref{66}. All modification is made in the egress port of the switch, which stands for the competing congestion of each output link. And we only use 3 (of 12) logical stages of a Tofino 1 programmable switch. 

The incoming packet's tenant ID (Tid) and timestamp (TS) are taken as input along with the forwarding process of the packet.
For the first stage, we use two registers to record information. The first register records the tenant ID of the last arriving packet and uses the second register for the storage of the timestamp of the last arriving packet (LTS). When a packet arrives, the value in these two registers will be replaced by the information in the packet and sent the old value to the next stage.
And then, in stage 2, the switch will judge two requirements: whether the Tid is the same as the last packet's tenant Id (LTid) also whether the time interval between LST and the TS is smaller than the timeout threshold (TH). We use the third register to record the timestamp of the last different tenant's packet's timestamp (LDTS). If both of these two requirements are satisfied, this register will update the value of LDTS with LTs and send the old LDTS to the next stage. Otherwise, just send LDTS value to the next stage.
%
For the last stage, the switch will simply compare the time interval between TS and LDTS to the TH. If the former is larger, then the inter-congestion will be determined, and the packet will be tagged and sent. If not, the packet will be sent without modification.

%

In this way, the programmable switch can easily identify the inter-tenant congestion situation and let the intended congested packets carry the signal to the receiver for our CAWC congestion control mechanism. Also, we can achieve this in a really simple implementation and resource usage of the switch, which can hardly influence the performance of \sys.

\section{Evaluation}
\label{sec7}

We evaluate \sys~both in large-scale NS-3 network simulation\cite{ns3} and testbed.
%
%
%
%
%
Firstly, we show that the weighted bandwidth allocation among tenants is achievable by using \sys. 
After that, we evaluate the bandwidth guarantee and work-conserving ability of \sys, using the ideal PS-N as the baseline of \sys.
\sys~reduces the packet loss ratio compared with HCSFQ, a state-of-the-art queue scheduling approach.
Hence, the flow completion time is reduced accordingly.
At last, we use the testbed to evaluate the transmitting lost and latency of the coordination job of \sys.

\noindent\textbf{Parameter Setup. } For \sys, we have a set of default settings. 
The update cycle of the byte-counter is the time period for \sys~to refresh and collect the bandwidth usage from each host for the allocation adjustment for the next period.
This cycle time decides the time granularity of \sys.
With a shorter value, more quickly can \sys~converge, at the same time bringing more communication costs between hosts and the coordinator. 
The cycle time we set in our experiments later is 0.01s. 
CAWC period stands for the frequency of congestion checking and feedback of \sys. 
The more this value is set, the faster will \sys~detect the congestion situation but come along with more traffic occupied by the feedback packets. 
In our evaluation, it is set to per 50 packets. 
Last but not least is the accelerated ratio of the rate controller for the allocation convergence. 
The higher this is set, the faster convergence might come. However, intenser shaking might also be caused. 
This value is set as $10\%$ in our tests.
The time-out value of the tenant-register-table in the switch is set as 0.1s in our evaluation which means if a unit-flow is not passing through the switch for 0.1s, it will be regarded as expired.
The lower this value is set, the preciser the congestion detection will be, as long as a heavier workload will occur on network traffic and switch.

\noindent\textbf{Metrics. }
We have four major performance metrics:
(\romannumeral1) Throughput, 
(\romannumeral2) packet dropping ratio,
(\romannumeral3) flow complete time (FCT), and 
(\romannumeral4) the fairness and accuracy of bandwidth allocation.

\begin{figure}[t]
  \centering

  \resizebox{\linewidth}{!}{
    \begin{subfigure}[t]{0.5\linewidth}
      \centering
      \resizebox{\textwidth}{!}{
        \includegraphics{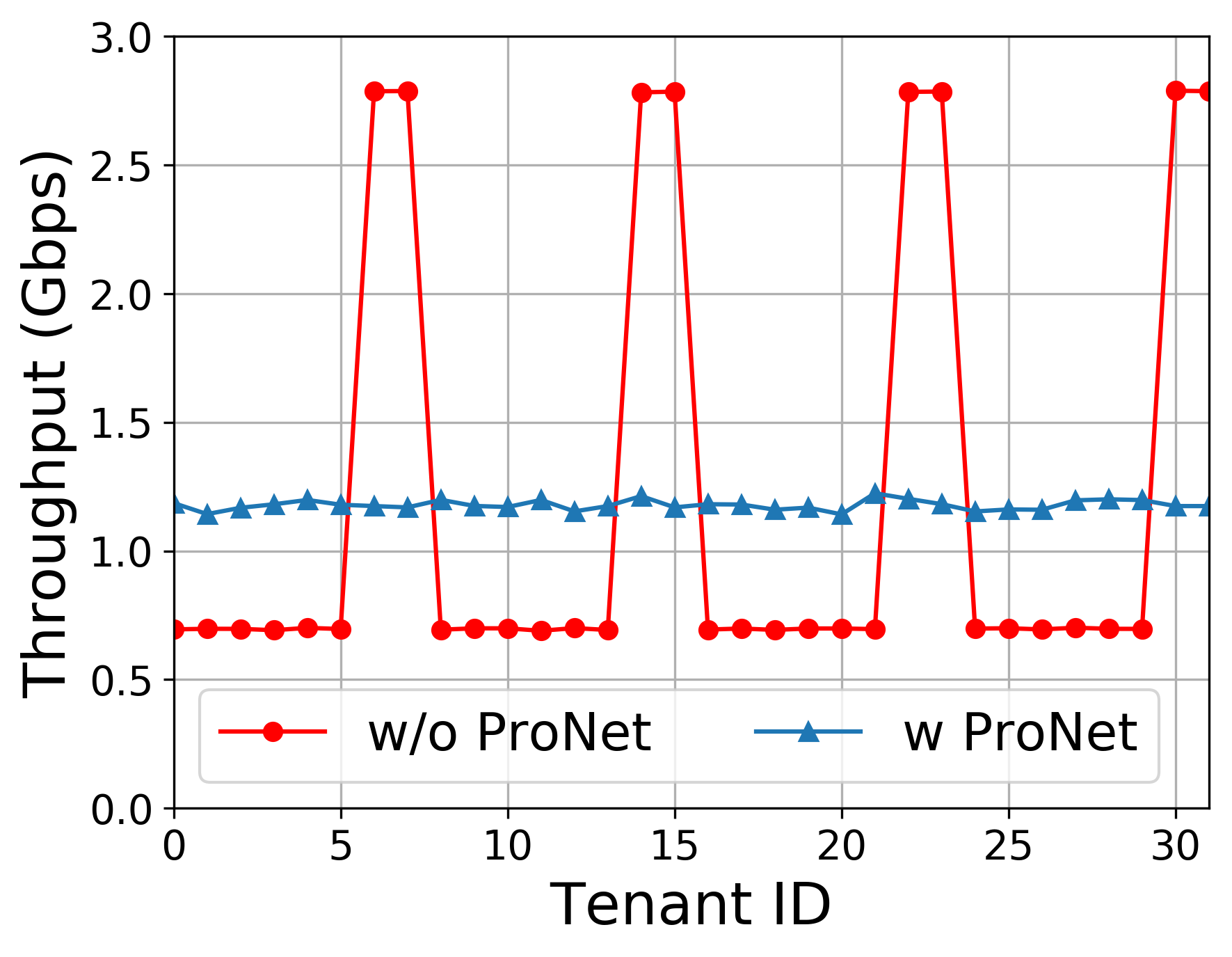}
      }
      \caption{Equal Weights}
      \label{7a}
    \end{subfigure}\hfil

    \begin{subfigure}[t]{0.5\linewidth}
      \centering
      \resizebox{\textwidth}{!}{
        \includegraphics{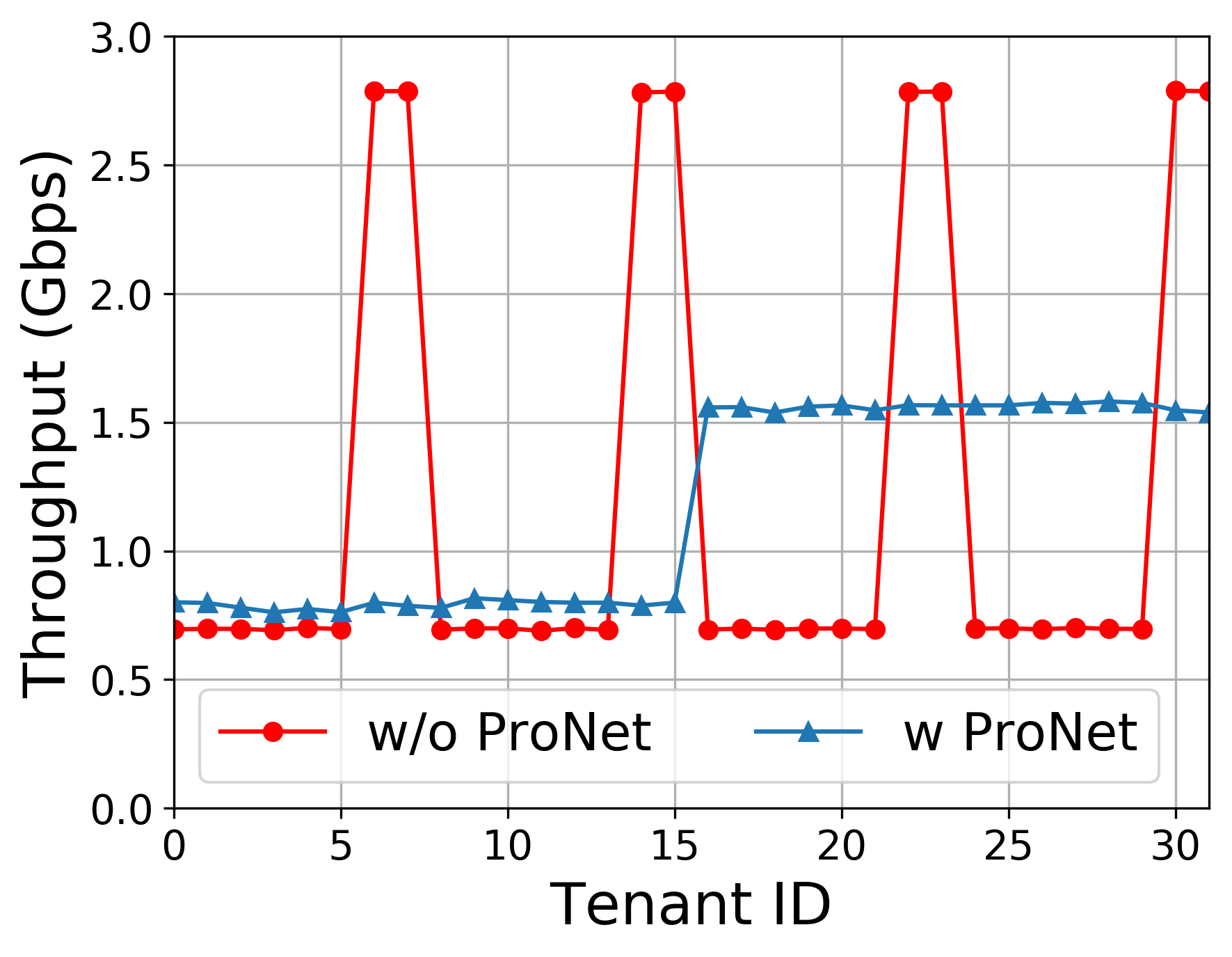}
      }
      \caption{Weighted allocation}
      \label{7b}
    \end{subfigure}\hfil

      }
    \setlength{\belowcaptionskip}{-3mm}
	\setlength{\abovecaptionskip}{2 mm}
  \caption{Allocation experiments with UDP traffic.}
  \label{f7}
\end{figure}

\begin{figure}[t]
  \centering
  \resizebox{\linewidth}{!}{
    \begin{subfigure}[t]{0.5\linewidth}
      \centering
      \resizebox{\textwidth}{!}{
        \includegraphics{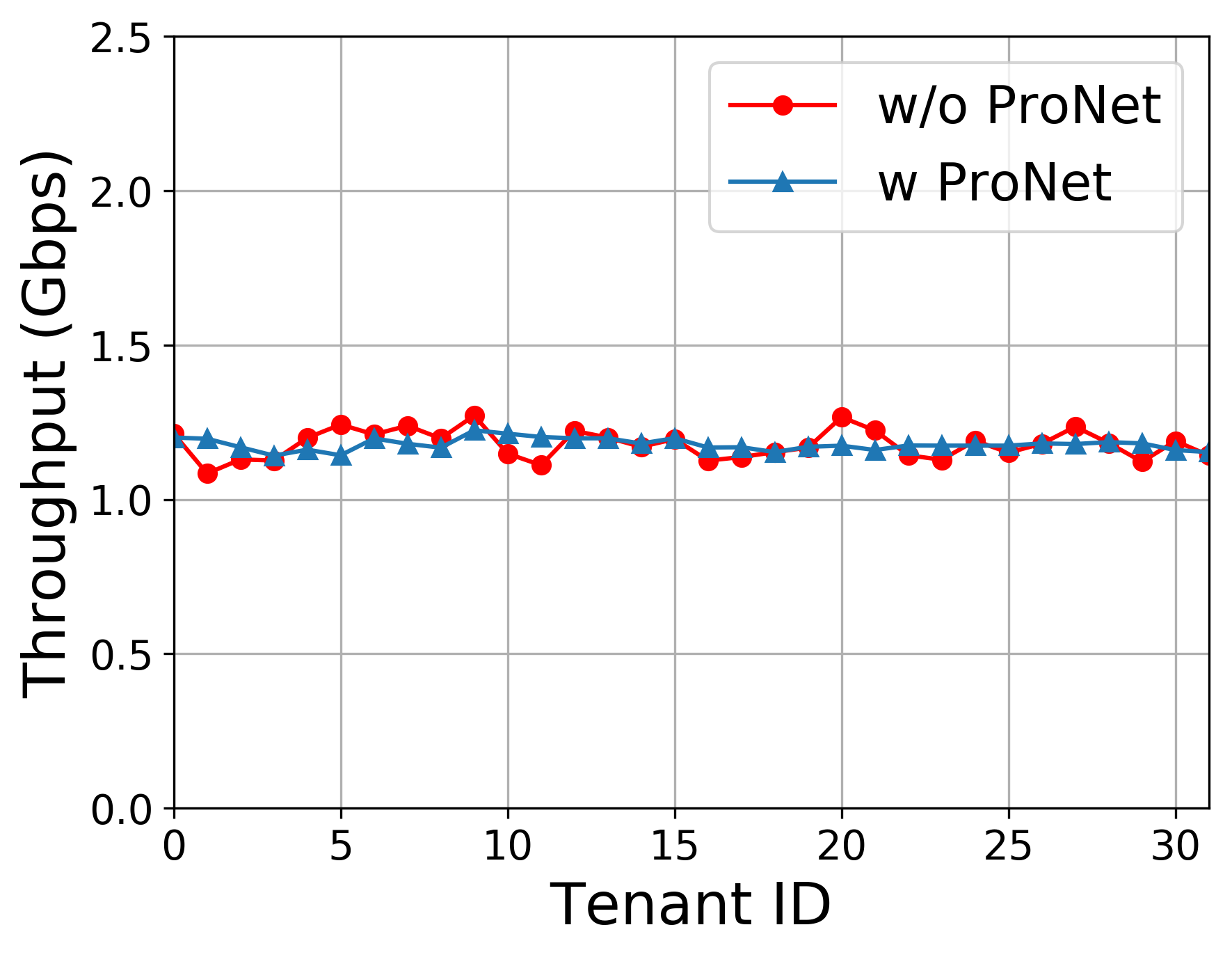}
      }
      \caption{Equal Weights}
      \label{8a}
    \end{subfigure}\hfil

    \begin{subfigure}[t]{0.5\linewidth}
      \centering
      \resizebox{\textwidth}{!}{
        \includegraphics{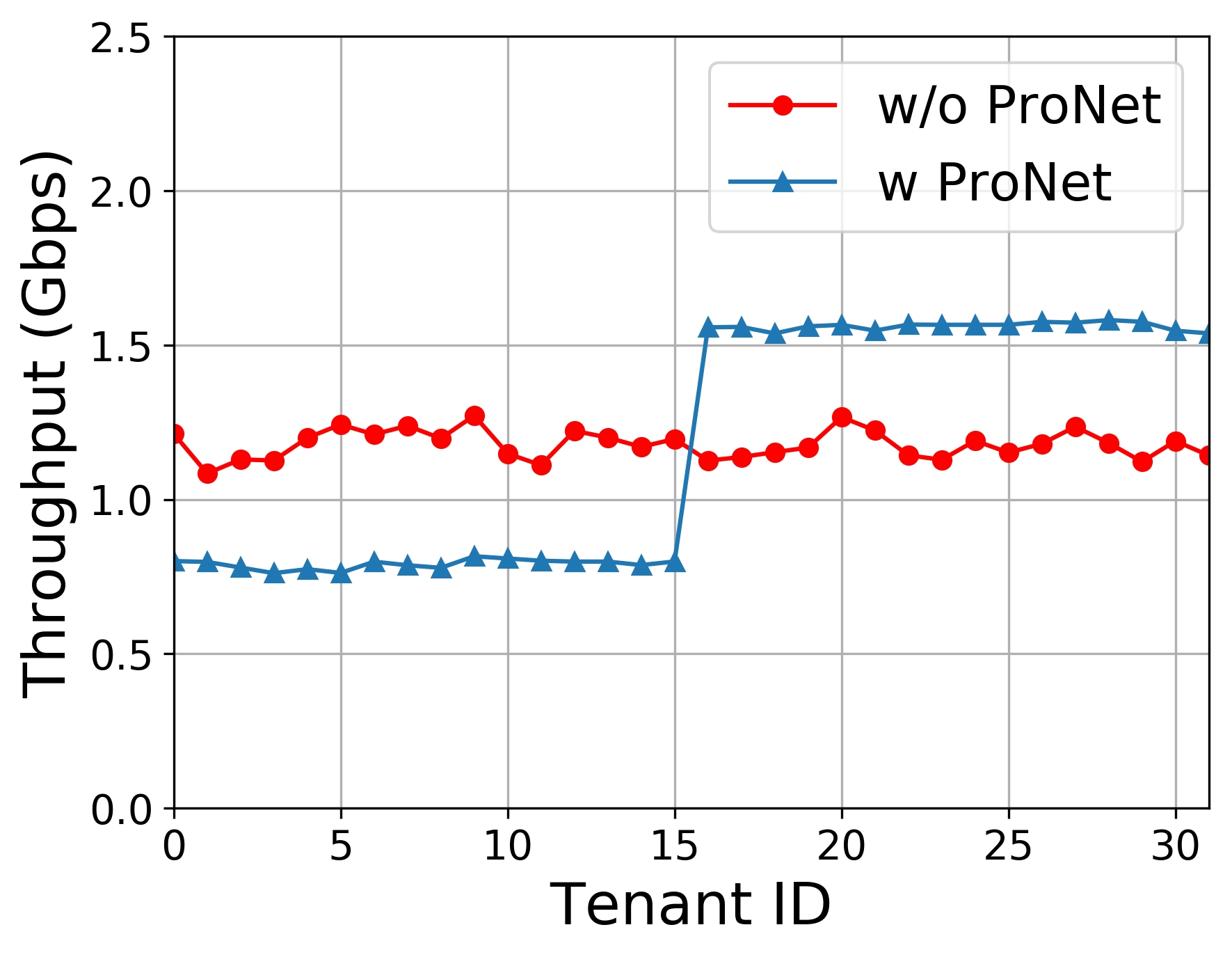}
      }
      \caption{Weighted allocation}
      \label{8b}
    \end{subfigure}\hfil

      }
    \setlength{\belowcaptionskip}{-3mm}
	\setlength{\abovecaptionskip}{2 mm}      
  \caption{Allocation experiments with TCP traffic.}
  \label{f8}
\end{figure}

\begin{figure}[t]
  \centering
  \resizebox{\linewidth}{!}{
    \begin{subfigure}[t]{0.5\linewidth}
      \centering
      \resizebox{\textwidth}{!}{
        \includegraphics{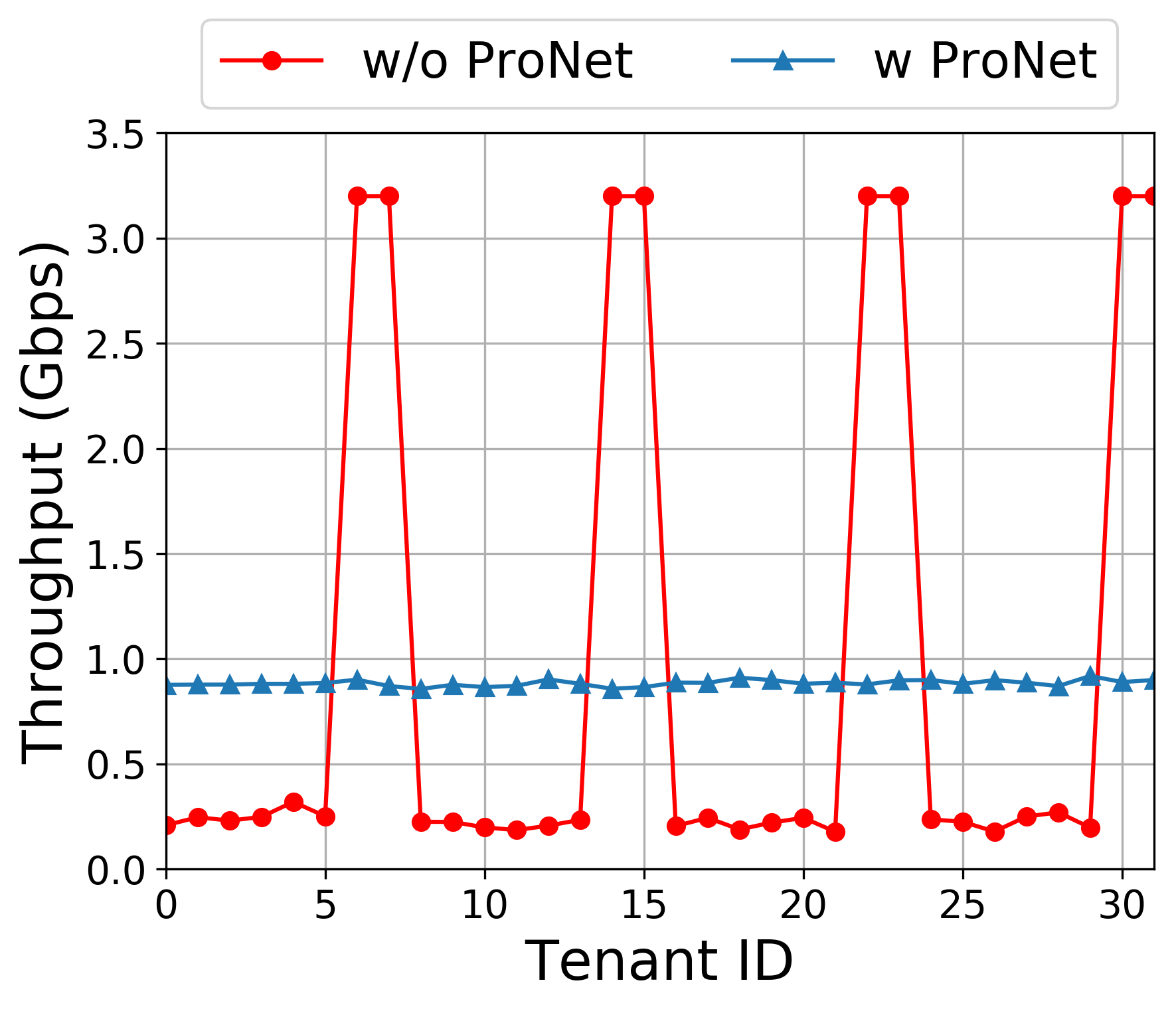}
      }
      \caption{Equal Weights}
      \label{9a}
    \end{subfigure}\hfil

    \begin{subfigure}[t]{0.5\linewidth}
      \centering
      \resizebox{\textwidth}{!}{
        \includegraphics{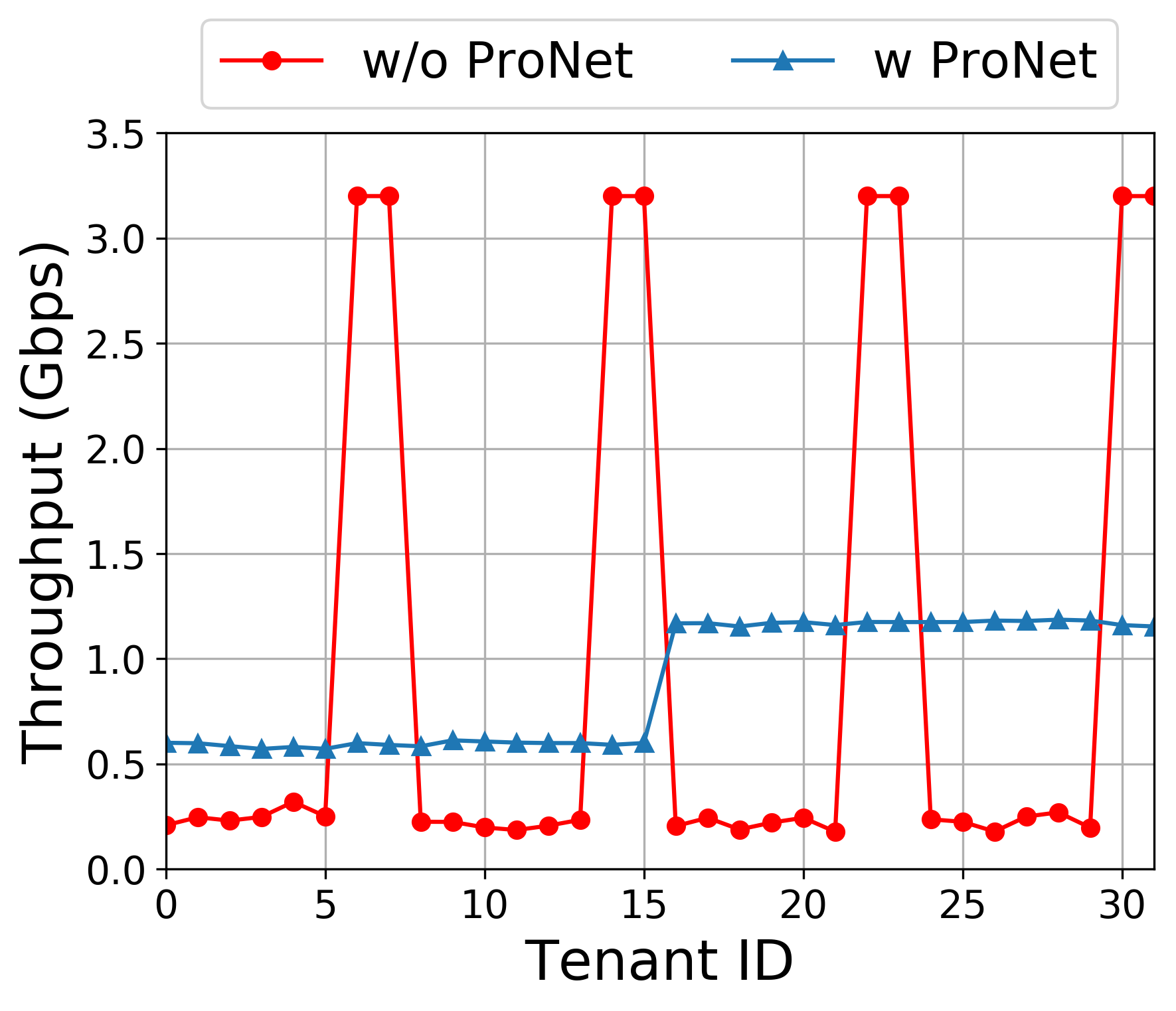}
      }
      \caption{Weighted allocation}
      \label{9b}
    \end{subfigure}\hfil

      }
          \setlength{\belowcaptionskip}{-3mm}
	\setlength{\abovecaptionskip}{2 mm}
  \caption{Allocation experiments with mixed traffic.}
  \label{f9}
\end{figure}

\subsection{Weighted Fairness Allocation Experiments}
Firstly, we prove that the weighted fairness allocation among tenants is guaranteed by using \sys. We cover both UDP and TCP traffic with equal or different weights. In the experiments, we use the topology based on Clos\cite{clos1953study}, which is a k=4 fat-tree topology. It has eight servers as the senders and two servers as the receiver, and we send packets in two groups. Each sender sends 8 flows (based on five-tuple), and a total of 64 flows are sent to receivers, and we select 32 flows for demonstration. All servers and switches are connected to 40Gbps links. Also, the bottleneck of our topology is 40Gbps. 




As the control group, we imply the traffic above with a normal UDP and TCP traffic with our \sys.
For the UDP tests, we set different rates for the UDP flows. 24 flows (Flow 1-24) are sent at 2Gbps, and 8 flows (Flow 25-32) are sent at 8Gbps. As shown in Figure\ref{f7}, flows 25-32 achieved a much higher bandwidth than flows 1-24, which does not meet the fairness requirement.
For the TCP tests, we also performed a simulation of TCP flows, and as shown in the Figure \ref{f8}, the 32 flows achieve a nearly fair sending rate due to the congestion control in the TCP protocol.
We also performed experiments for the mixed TCP and UDP flow situation. As shown in the figure, flows 6, 7, 14, 15, 22, 23, 30, and 31 are UDP flows, and others are TCP lows sent at 3.2Gbps. As shown in the Figure \ref{f9}, without ProNet, the congestion control in TCP protocol cannot interfere with the UDP flows. Thus the UDP flow sending rate is much higher than TCP flows, which does not meet the requirement of fairness in multi-tenant networks.

\noindent\textbf{\sys can guarantee the fairness allocation among tenants. } As for the performance of \sys, first of all, 
in order to show the ability of \sys~to keep fairness, we set all of the tenants in the same weights. In Figure \ref{7a}, \ref{8a}, and \ref{9a}, although they have the same weight, we set the initial sending rates differently. We set 16 flows with the same weight. Flows 1-16 belong to tenant 1 and flows 17-32 belong to tenant 2. 


As shown in Figure \ref{f7}-\ref{f9}, for all of these tests, using \sys, we can see that the fairness among flows is kept at the same level, regardless of the network protocol and communication pattern of the network. These experiments show the ability of \sys to achieve the fairness allocation among tenants and flows.

\noindent\textbf{\sys can guarantee the weighed fairness allocation among tenants. } The weighted allocation among tenants is also shown in our experiments. In our setups, flows 1-16 belong to tenant 1, and flows 17-32 belong to tenant 2. The relative weight between tenants 1 and 2 is assigned as 1:2. We also evaluate our system in UDP traffic, TCP traffic, and also the mix Traffic of them. Figure \ref{7b}, \ref{8b}, and \ref{9b} shows the results. We can see that the flows in two different tenants are properly allocated the bandwidth as the preset ratio. All flows in tenant 1 keep a throughput of about 1/2 of the flows in tenant 2 for the whole 40Gbps bandwidth. Especially for the TCP and UDP mix traffic pattern, we have random flows of UDP and TCP from each tenant's traffic. In Figure \ref{f9}, our system can still perform well for the weighted allocation between these two tenants as a 1:2 ratio.

\subsection{Work Conservation and Bandwidth Guarantee Experiments}

\begin{figure}[t]
  \centering
  \resizebox{\linewidth}{!}{
    \begin{subfigure}[t]{0.5\linewidth}
      \centering
      \resizebox{\textwidth}{!}{
        \includegraphics{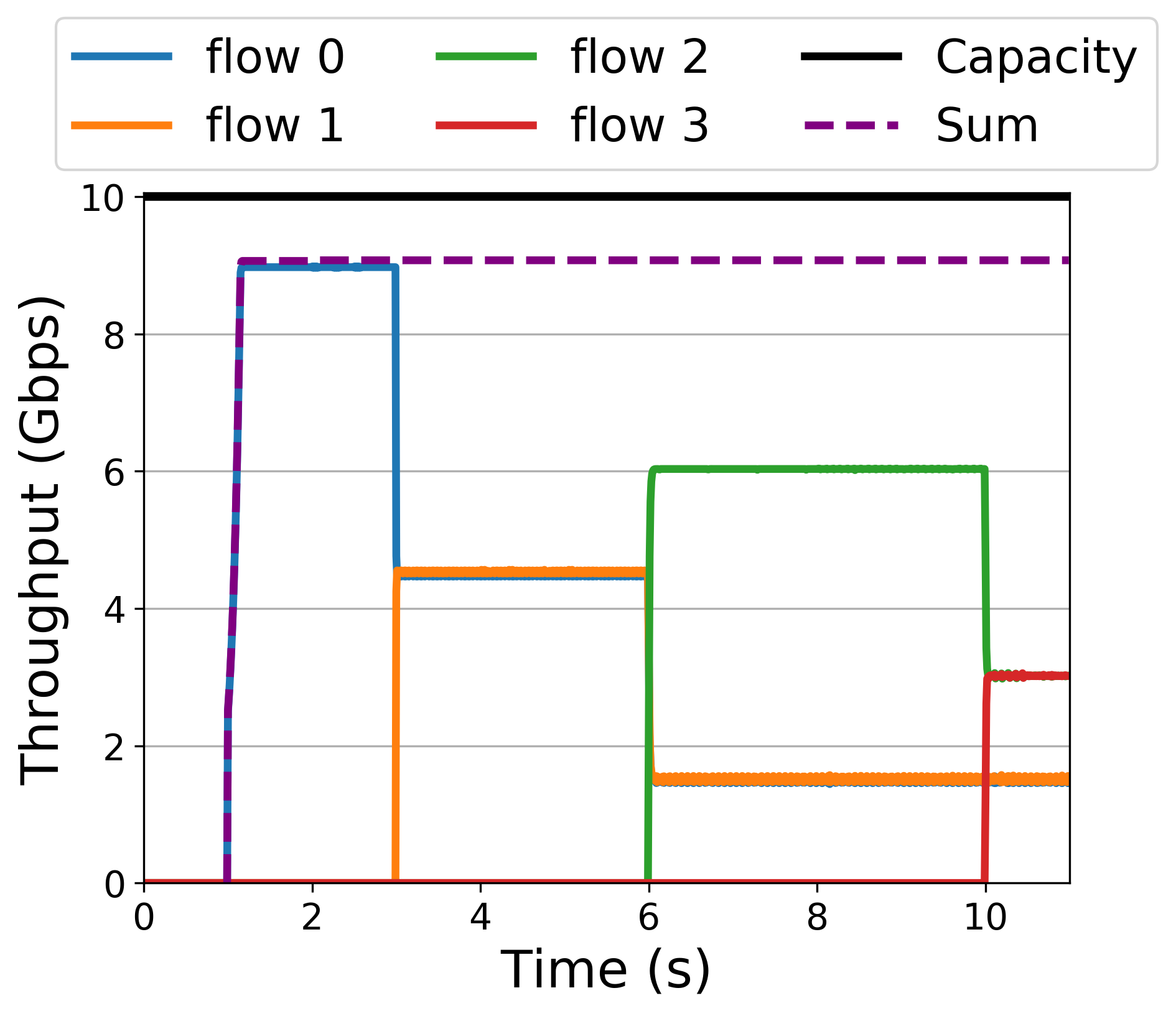}
      }
      \caption{Work conservation Test \protect\\ for the full usage of the 10Gb\protect\\ total bandwidth.}
      \label{10a}
    \end{subfigure}\hfil

    \begin{subfigure}[t]{0.5\linewidth}
      \centering
      \resizebox{\textwidth}{!}{
        \includegraphics{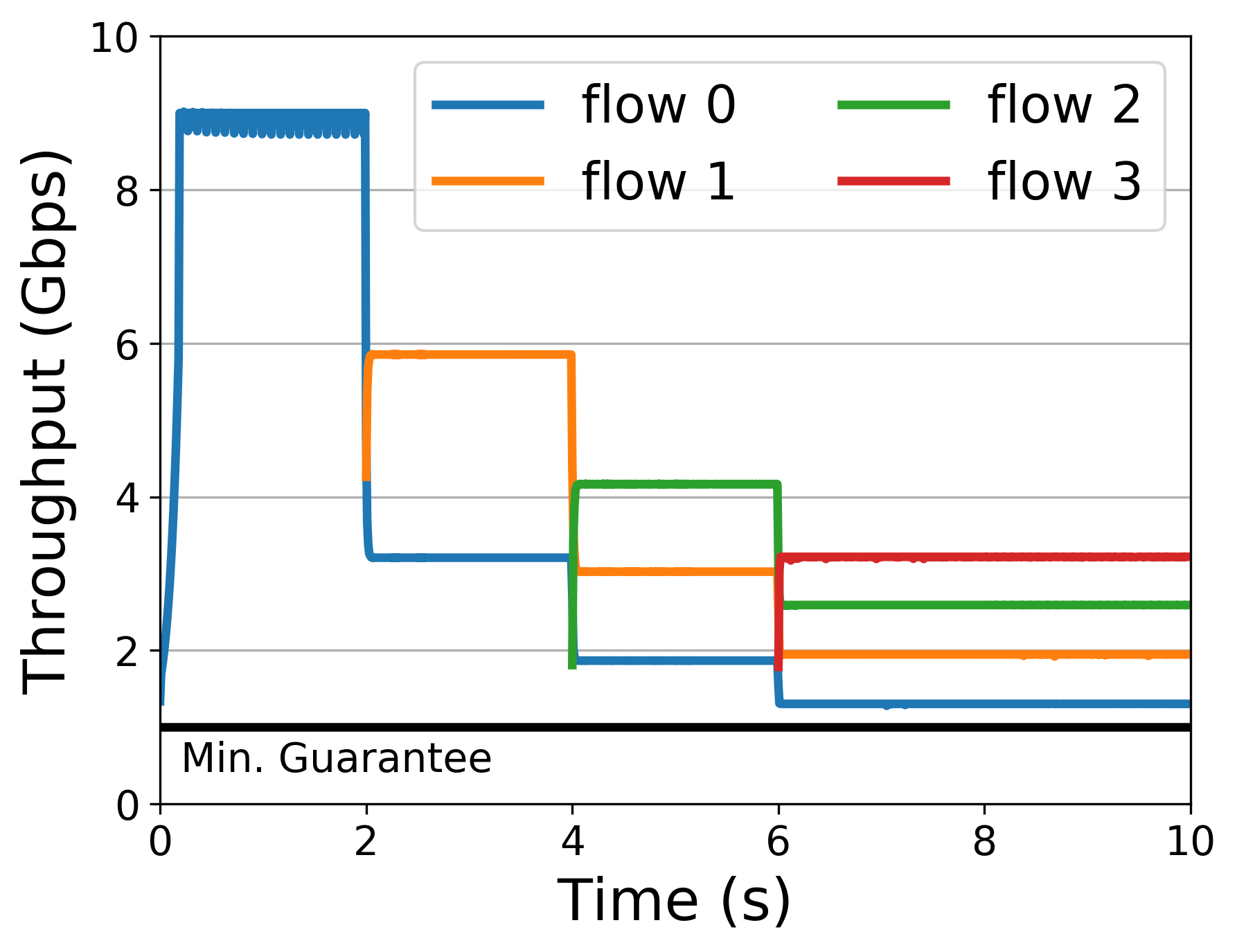}
      }
      \caption{A bandwidth guarantee experiment as a min. bandwidth set as 1Gbps.}
      \label{10b}
    \end{subfigure}\hfil

      }
          \setlength{\belowcaptionskip}{-5mm}
	\setlength{\abovecaptionskip}{2 mm}
  \caption{Work conservation and bandwidth guarantee experiments.}
  \label{f10}
\end{figure}




\noindent\textbf{Work Conservation Experiment. }
 For the work conservation part, we meanly monitor the congestion links, which is the critical allocation path in the whole network. In this experiment, we set 4 flows from 2 different tenants with different arrive and leave times. The tenants' weight ratio is assigned as 1:2, flow 0 and flow 1 belong to tenant 1, and the other two belong to tenant 2. All these four flows share the total 10Gbps bottleneck. The weights assigned to the flows inside these tenants are equal. And for the traffic, we mainly concentrate on the congested link to track the behavior of ProNet. The result is shown in Figure \ref{10a}. 
 
As a comparison, we also calculate the summary of these flows' bandwidth usage shown in the figure, also with the full bandwidth capacity. With the result in Figure \ref{10a}, in the congestion link, each flow can immediately acquire the correct bandwidth share, and at the same time, all flows in the bottleneck link can occupy the total capacity of the link. For example, at 6s, flow 2 comes into the bottleneck. The allocation of these three flows here instantly changes. As shown in the figure, flow 2 occupies about 6 Gbps bandwidth, and for the other two flows from another tenant, flows 0 and 1 are adjusted to sharing an amount of about 3 Gbps bandwidth equally. This is a perfect example of the correctness of \sys. Above all, this test shows our system can achieve the work conservation whenever flows come or leave. 

\noindent\textbf{Bandwidth Guarantee Experiment. }
We also evaluate the minimum bandwidth guarantee of \sys. In this experiment, we set several flows from different tenants with different allocation weights. Also, for each tenant's flow, have set a minimum bandwidth for the allocation. And with the minimum guarantee preset by using the bandwidth function and other mechanisms in \sys~. In the experiment, flows 1 to 4 are from different tenants, and the ratio between these four tenants is 1:2:3:4. As for the minimum bandwidth, all these tenants are set as a 10Gbps bandwidth guarantee by setting the corresponding bandwidth functions. Figure \ref{10b} shows the testing result. Also, with the min. guarantee marked in the figure, we can see that all the throughput of each tenant' flows starts all at their guaranteed throughput regardless of the condition of other flows. Also, all flows can achieve the preset weights bandwidth allocation ratio. This experiment shows the minimum guarantee goal can be achieved by using \sys.


\begin{figure}[t]
  \centering
  \resizebox{\linewidth}{!}{
    \begin{subfigure}[t]{0.5\linewidth}
      \centering
      \resizebox{\textwidth}{!}{
        \includegraphics{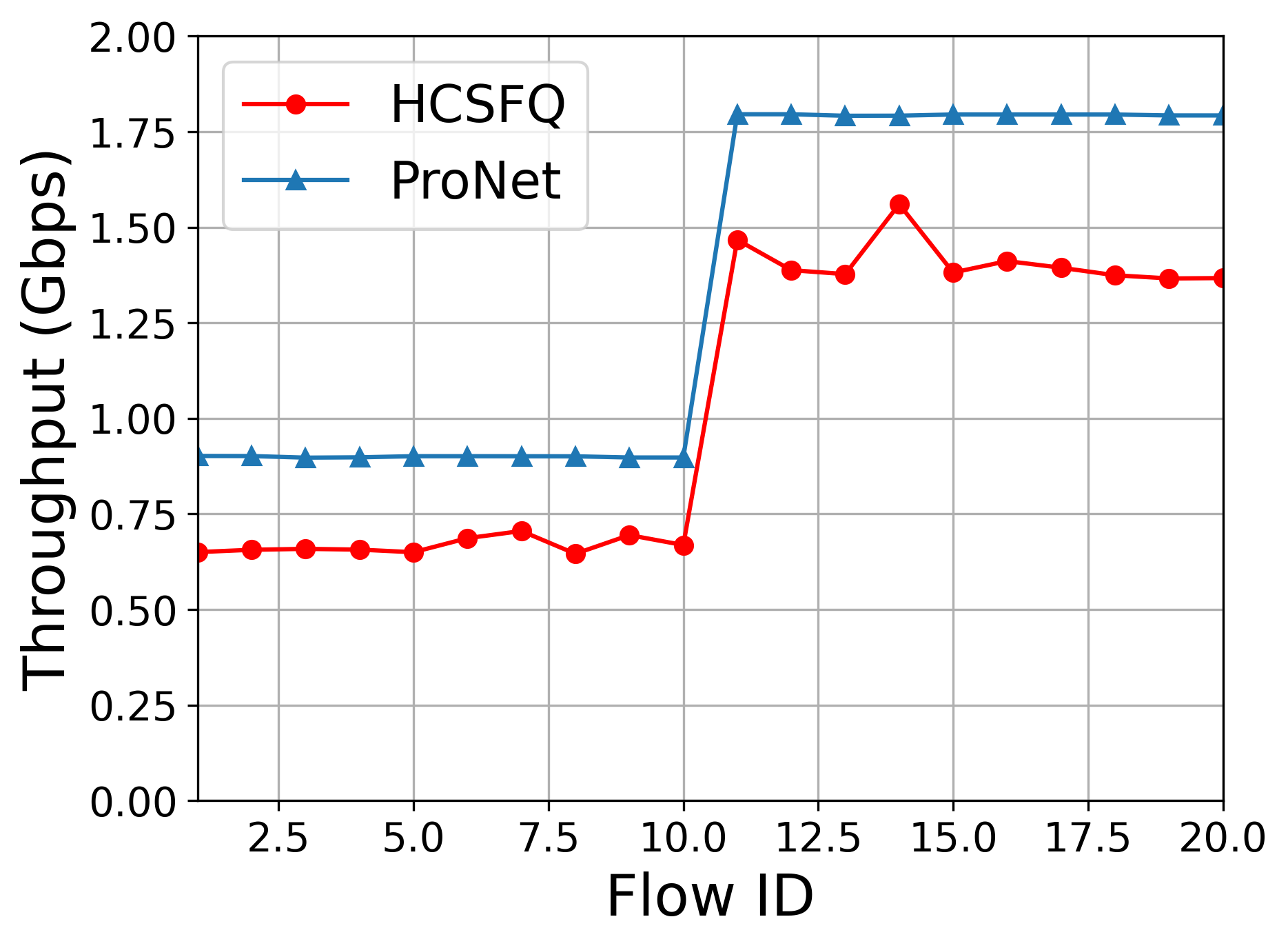}
      }
      \caption{TCP NewReno}
      \label{fig:testbed-topos:a}
    \end{subfigure}\hfil

    \begin{subfigure}[t]{0.5\linewidth}
      \centering
      \resizebox{\textwidth}{!}{
        \includegraphics{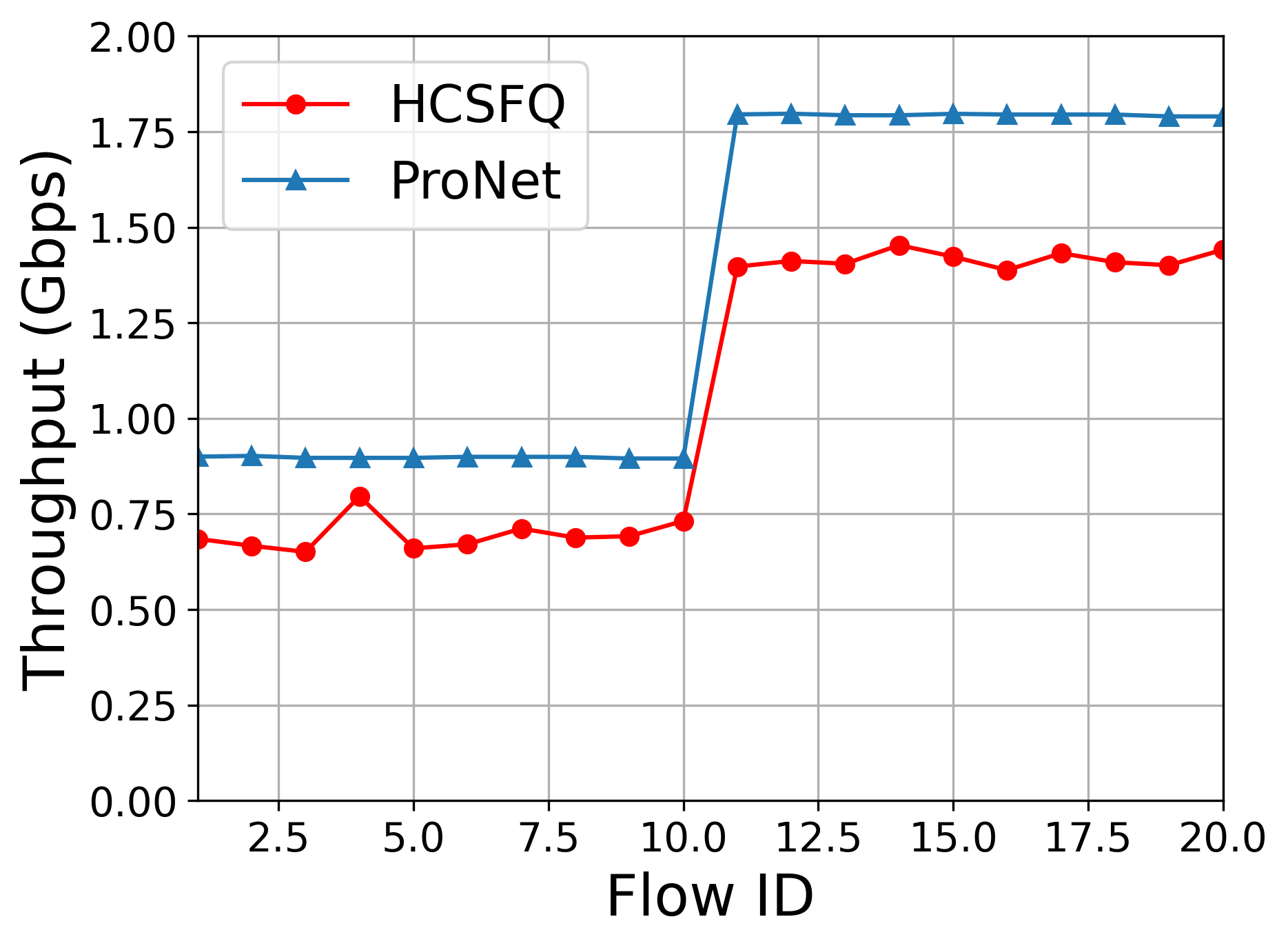}
      }
      \caption{TCP BIC}
      \label{fig:testbed-topos:b}
    \end{subfigure}\hfil

      }
    \setlength{\belowcaptionskip}{-5mm}
	\setlength{\abovecaptionskip}{2 mm}
  \caption{Throughput of flows between HCSFQ and \sys under different TCP protocols.}
  \label{fig:testbed-topos}
\end{figure}

\subsection{Performance comparison with HCSFQ}\label{subsec:pqt}
As shown in Figure \ref{fig:testbed-topos}, we compare \sys with HCSFQ, a state-of-art work that implements weight-fair-queuing through active packet dropping in programmable switches. 
Results show that \sys~achieve better performance than HCSFQ in weighted share allocation.
%
%
We measure the performance of \sys~and HCSFQ on a simple topology, where three hosts, A, B, and C, are connected by 30 Gbps, 1 \textmu s-delay cables to a single switch with a maximum buffer size of 250KB per port.
Hosts A and B each start 10 TCP flows, which send 0.2GB and 0.1GB to host C, respectively. 
The TCP flows started by host A have twice the weight of that of host B.
ECN is enabled in this experiment, with a minimum threshold of 50KB and a maximum threshold of 200KB.
To avoid pathological behavior of TCP flows under HCSFQ's proactive packet dropping, both the reconnect time-out (time-out after an SYN packet is not responded) and the minimum RTO are set to 10 ms.
As shown in Figure \ref{fig:testbed-topos:a}, the throughput of the flows with \sys is always higher than the throughput of flows with HCSFQ under TCP NewReno.
This is because HCSFQ's proactive packet dropping wastes network bandwidth on the one hand and causes the congestion window to be too low to utilize the bandwidth on the other.
In the experiment, HCSFQ proactively drops about 5\% of the packets, which resulted in a severe performance impairment of the TCP flows despite a small time-out set to alleviate the impairment of packets dropping.
In contrast, flows with \sys have almost no packet loss because the rate is precisely controlled.
As a result, the total throughput of flows with HCSFQ is only 78\% of that of flows with \sys.
For FCT, HCSFQ is on average 31\% longer than \sys.
Another noticeable observation is that \sys provides fairer bandwidth allocation than HCSFQ, with the \textit{coefficient of variation}
\footnote{The \textit{coefficient of variation} is also known as \textit{relative standard deviation}, which is defined as the ratio of the \textit{standard deviation} to the \textit{mean}. The higher the \textit{coefficient of variation}, the greater the dispersion.}
of throughput of flows with HCSFQ is 36 times higher than that with \sys.
Finally, it is found that \sys can accurately allocate bandwidth in proportion to the weight, with a throughput ratio of $1.99$ for two groups of flows in \sys and $2.11$ for that in HCSFQ.
As demonstrated in Figure \ref{fig:testbed-topos:b}, we also compare \sys with HCSFQ under other congestion control protocols, and the aforementioned results still hold, with \sys outperforming HCSFQ in terms of throughput, fairness, and accuracy of bandwidth allocation.
%
%
%
%
%

\begin{figure}[t]
  \centering
  \resizebox{\linewidth}{!}{
    \begin{subfigure}[t]{0.5\linewidth}
      \centering
   
        \includegraphics[width=0.95\linewidth]{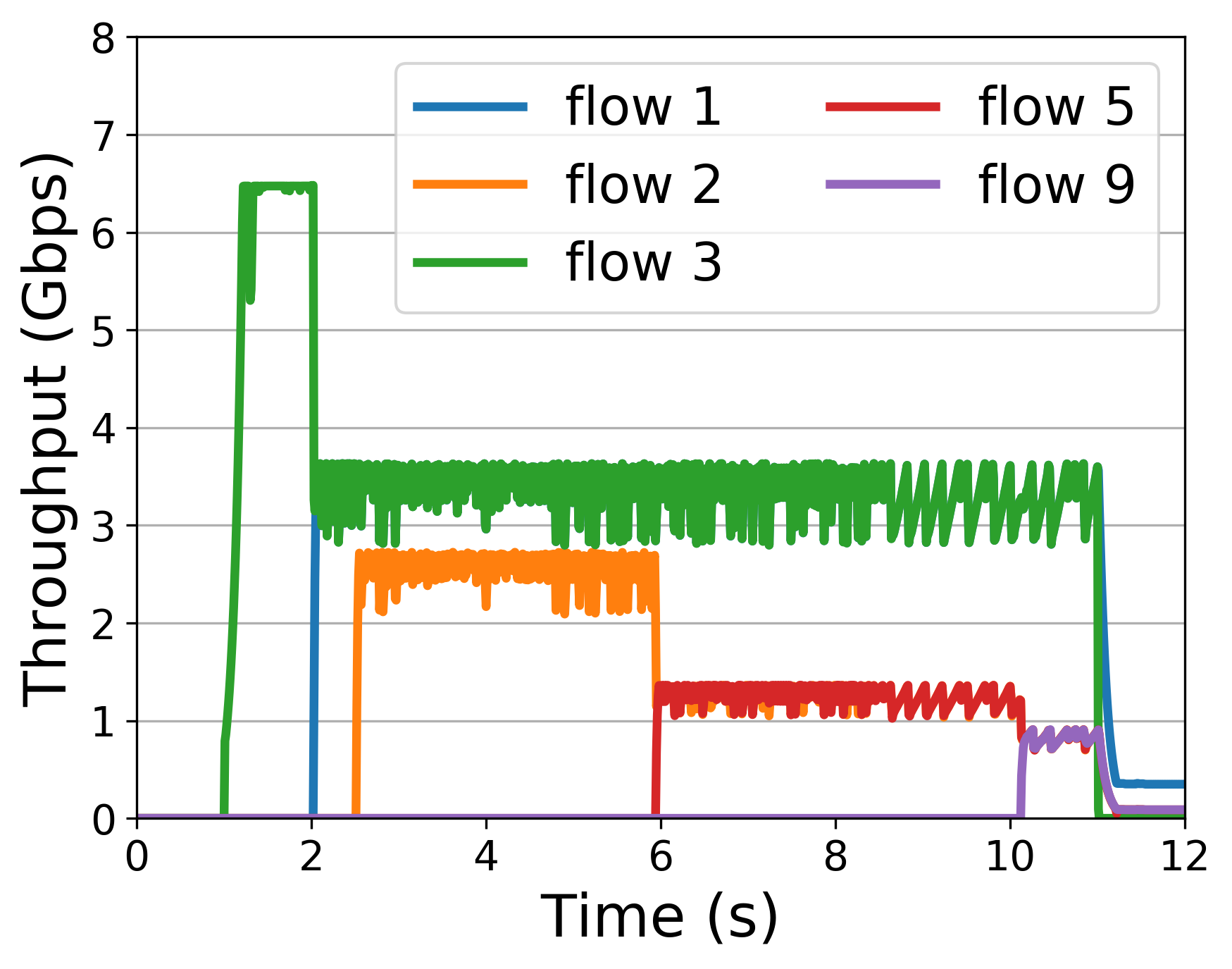}

      \caption{Flows' bandwidth result.}
      \label{12a}
    \end{subfigure}\hfil

    \begin{subfigure}[t]{0.5\linewidth}
      \centering
        \includegraphics[width=0.95\linewidth]{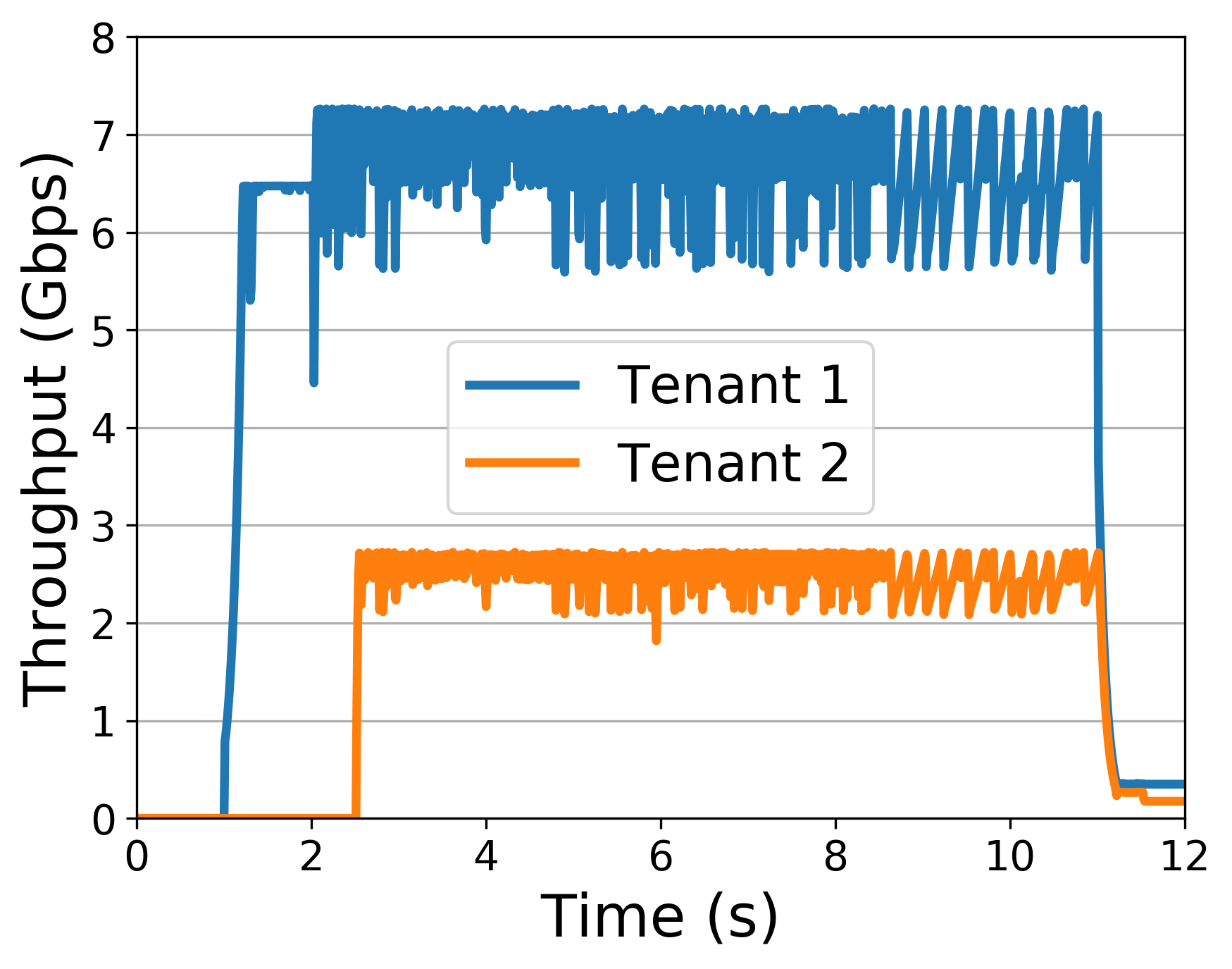}
      
      \caption{Tenants' bandwidth result.}
      \label{12b}
    \end{subfigure}\hfil

      }
          \setlength{\belowcaptionskip}{-5mm}
	\setlength{\abovecaptionskip}{2 mm}
  \caption{Large scale experiments result.}
  \label{f12}
\end{figure}



\subsection{Large Scale Experiments}
In this part, we evaluate the stability and the correctness of the large-scale simulation experiments. The topology we use is a k=10 which contains 250 hosts and tens of switching nodes in it to simulate the real network condition. We have deployed over 2000 Poisson random flows belonging to 48 tenants with different weight ratios. The flows we set are to simulate the web searching task, which is commonly seen in the datacenter, containing burst flows, short flows, and other kinds of uncommon flows that might appear in a datacenter. And the experiment is deployed for a relatively long period of time to prove the overall network allocation situation. 

\noindent\textbf{\sys can guarantee the fairness allocation among tenants and flows in large-scale experiments. } To show the allocation result of ProNet more clearly, we choose some representative flows and tenants to show in our paper. Figure \ref{f12} shows the result of this allocation experiment. 
In figure \ref{12a}, we pick 5 flows, flows 1 and 3 belong to a tenant with a normalized weight of 2, and other flows are from another tenant with a normalized weight of 1. We can see that all the tenants (' flows) share can be kept relative stability and remains consistent with the same normalized share value. Also, figure \ref{12b} shows the allocation result in a bottleneck link between two tenants. The tenants we choose in this figure have a 2:1 weight ratio. And we can see the allocation among flows is also fairness guaranteed. As a result, tenants 0-23 have an average throughput of around 1.239 Gbps, and tenants 24-27 have an average throughput of around 2.423 Gbps.
Tenants 1 to 24 have normalized weight 1, and others have normalized weight 2. We can see in the result, \sys can achieve the weighted bandwidth allocation perfectly among tenants. All in all, we can see \sys~is able to perform well in the real datacenter environment and is a practical bandwidth management system.

\subsection{Congestion Awareness Experiments}
In this part, our experiments are on our real-machine testbed with a Tofino 1 switch for the evaluation of our tenant-counter design with the implementation we mentioned in \S~\ref{sec6}. And to prove the ability to detect different kinds of congestion we mentioned in \S~\ref{sec:system:tenant-counter}. The topology is set as two links between two servers. One of the links has a programmable switch installed in the middle, and the other link is installed with one ordinary switch. Here, we meanly prove the tenant-counter mentioned above to detect the intra-tenant congestion by using the programmable switch is feasible and deployable. 

\begin{figure}[t]
  \centering
  \resizebox{\linewidth}{!}{
    \begin{subfigure}[t]{0.5\linewidth}
      \centering
      \resizebox{\textwidth}{!}{
        \includegraphics{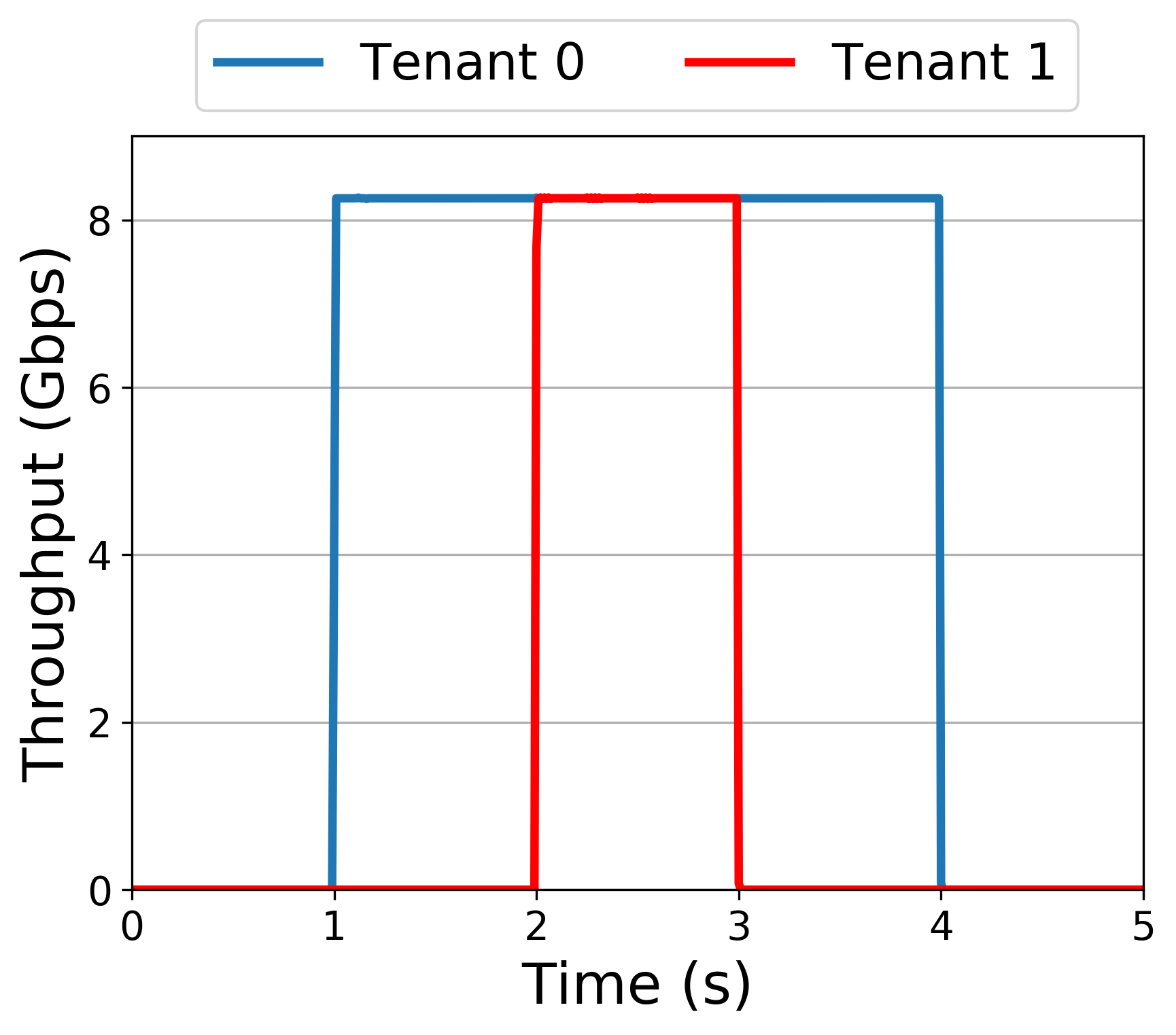}
      }
      \caption{Intra-tenant Congestion.}
      \label{13a}
    \end{subfigure}\hfil

    \begin{subfigure}[t]{0.5\linewidth}
      \centering
      \resizebox{\textwidth}{!}{
        \includegraphics{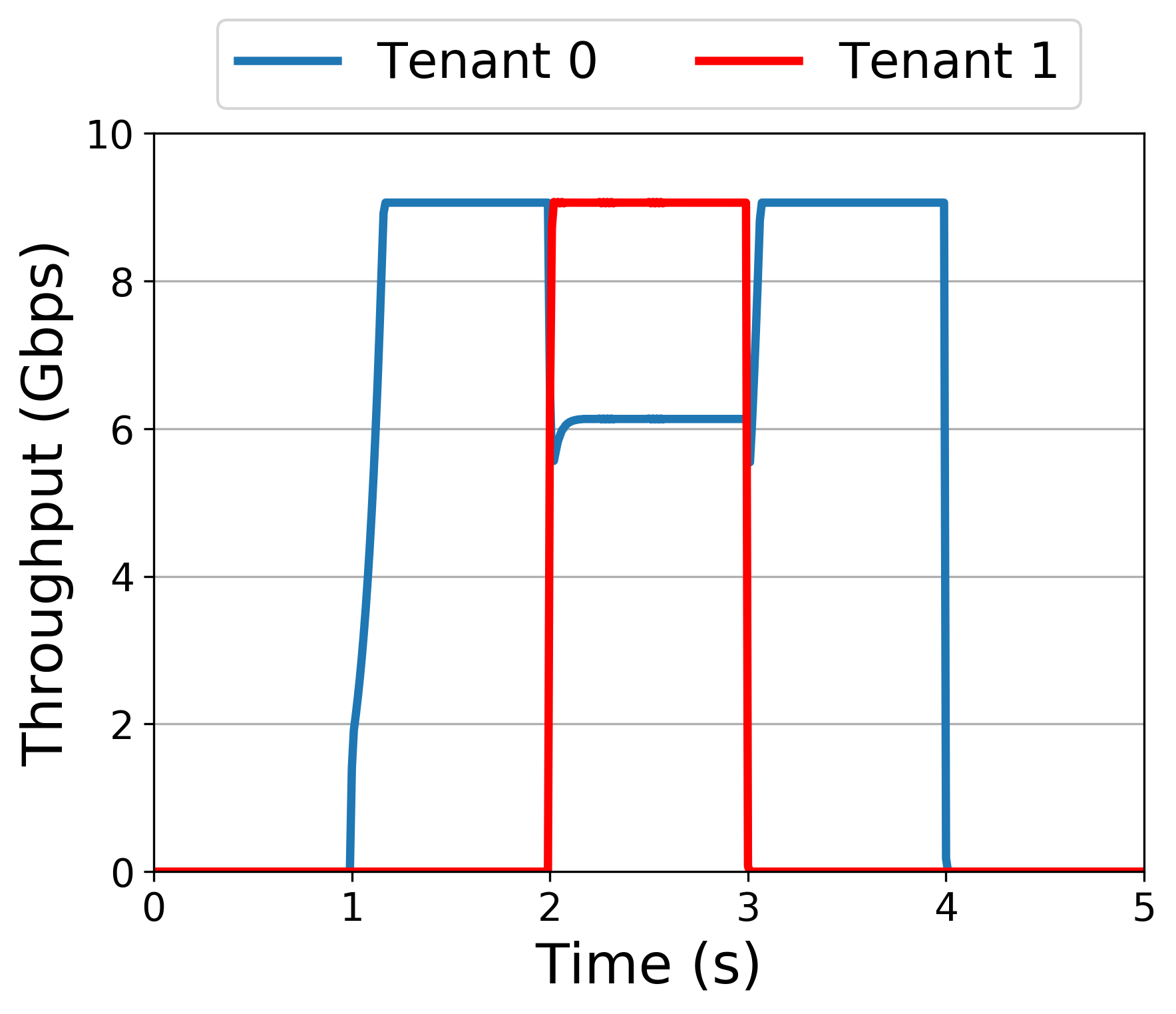}
      }
      \caption{Inter-tenant Congestion.}
      \label{13b}
    \end{subfigure}\hfil

      }
          \setlength{\belowcaptionskip}{-5mm}
	\setlength{\abovecaptionskip}{2 mm}      
  \caption{\sys~with tenant-counter Experiments.}
  \label{f13}
\end{figure}

We set up two scenarios in this experiment. In the first one, one of the links is used by flows all from tenant 1, and the other one is used by tenant 2, in which we create an intra-tenant congestion situation. For the second scenario, the flows in each link are mixed with flows from both tenants 1 and 2, which is a normal inter-tenant congestion situation. For the first one, the congestion signal is due to the flows in the same tenant, which shouldn't influence the allocation to the flows of other tenants. And the second one's congestion is caused by flows belonging to different tenants. Figure \ref{f13} shows the testing results. Figure \ref{13a} shows the intra-congestion situation, the allocation of other tenants' flow is not affected by it. However, in Figure \ref{13b}, the flows from different tenants are congested, and as shown in the result, the allocation performs as usual with a weighted share. 

This experiment proves the correctness and the feasibility of our design of tenant-counter in the deployment of the actual programmable switch.

%
%
%



\section{Conclusion}
\label{sec8}

In this paper, we propose \sys, a practical end-host-based bandwidth allocation protocol designed for private datacenters.
%
%
At the core of \sys, it leverages byte-counter to monitor and adjust the bandwidth usage on end-hosts.
\sys~supports proportional bandwidth allocation among tenants, and minimum bandwidth guarantee, simultaneously achieving work conservation.
In addition, flexible bandwidth allocation is also supported according to tenant-specified bandwidth functions.
%
%
\sys~improves the application-level performance by reducing the packet loss ratio and improving the network throughput.
Both our implementation and simulation results indicate that \sys~is a promising bandwidth allocation protocol.
\bibliographystyle{ACM-Reference-Format}
\bibliography{sample-base}

\appendix
\appendix
\section{Tenants' Bandwdith Function Aggregation}
\label{appendix:a}
\setcounter{table}{0}
\renewcommand{\thetable}{\Alph{section}\arabic{table}}

In this section, we introduce the progress and algorithm for the aggregation of BF which mainly inspired by the BwE paper.

The target (aggregated) bandwidth function $B_f^t\left(s\right)$ for unit-flow $f$ of tenant $t$ must satisfy Equation \eqref{2}, where $s$ denotes the fair share and $B_t\left(s\right)$ is the original (before aggregation) bandwidth function of the tenant. 
It ensures that bandwidth allocated to a tenant will be allocated to all of its unit-flows eventually, \ie, the sum of all unit-flows' aggregated BF should be equal to the original tenant's BF.

\begin{equation}\label{2}
\forall s,\ \sum_{\forall f|f\in t}{B_f^t\left(s\right)=}B_t\left(s\right)
\end{equation}

%
To satisfy Equation \eqref{2}, we first define an add-up bandwidth function $B_t^a$ by summing up the original bandwidth functions of all unit-flows of tenant $t$:

\begin{equation}\label{3}
\forall s,\ B_t^a\left(s\right)=\ \sum_{f\in t} B_f\left(s\right)
\end{equation}
Next, in order to link the original unit-flows' BF, which is represented by the add-up BF $B_t^a$ in \eqref{3} to the tenant's BF $B_t$, a transforming function $T$ from unit-flows to tenants that satisfies Equation \eqref{4} should be found:

\begin{equation}\label{4}
T\left(s\right)=\ s^{\prime}|B_t^a\left(s\right)=\ B_t\left(s^{\prime}\right)
\end{equation}

The transforming function $T$ in \sys is a mapping between the fair share $s$ of the add-up BF and the fair share $s^{\prime}$ of the tenant's which correspond to the same bandwidth value. 
%
%

%
At last, for each unit-flow $f\in t$, apply $T$ on $ B_f$'s fair share to get $B_f^e$ for each flow:

\begin{equation}\label{5}
B_f^t\left(T\left(s\right)\right)=B_t\left(s\right)
\end{equation}

In this way, \sys~can get aggregated BFs for each unit-flow which can both satisfy the BF of the unit-flow and the tenant it belongs to. That's essential for \sys to coordinate between tenants and their network flows.
%

%









\end{document}